\documentclass[aps,showpacs,nofootinbib,superscriptaddress]{revtex4-1}
\usepackage{epsf}
\usepackage{graphicx}
\usepackage{amsmath}
\def\slashchar#1{\setbox0=\hbox{$#1$}
   \dimen0=\wd0 \setbox1=\hbox{/} \dimen1=\wd1
   \ifdim\dimen0>\dimen1 \rlap{\hbox to \dimen0{\hfil/\hfil}} #1
   \else  \rlap{\hbox to \dimen1{\hfil$#1$\hfil}} / \fi}

\newcommand{\be}{\begin{equation}}
\newcommand{\ee}{\end{equation}}
\newcommand{\bea}{\begin{eqnarray}}
\newcommand{\eea}{\end{eqnarray}}

\def\g#1{\gamma_{#1}}
\makeatletter
\def\slashchar#1{{\mathpalette\c@ncel{#1}}} 
\makeatother
\def\vsl{\slashchar{v}}
\begin{document}

\title{ Exclusive $c\to s,d$ semileptonic decays of ground-state
spin-1/2 doubly charmed baryons }

\author{ C.Albertus} \affiliation{Departamento de F\'\i sica Fundamental e
IUFFyM,\\ Universidad de Salamanca, E-37008 Salamanca, Spain} \author{
E. Hern\'andez} \affiliation{Departamento de F\'\i sica Fundamental e
IUFFyM,\\ Universidad de Salamanca, E-37008 Salamanca, Spain}
\author{J.~Nieves} \affiliation{Instituto de F\'\i sica Corpuscular
(IFIC), Centro Mixto CSIC-Universidad de Valencia, Institutos de
Investigaci\'on de Paterna, Aptd. 22085, E-46071 Valencia, Spain}


\begin{abstract}
  We evaluate exclusive semileptonic decays of ground-state spin-1/2
  doubly heavy charmed baryons driven by a $c\to s,d$ transition at
  the quark level.  Our results for the form factors are consistent
  with heavy quark spin symmetry constraints which are valid in the
  limit of an infinitely massive charm quark and near zero recoil.
  Only a few exclusive semileptonic decay channels have been
  theoretically analyzed before. For those cases we find that our
  results are in a reasonable agreement with previous calculations.
\end{abstract}

\maketitle
%
%
%
%
%
%
\section{Introduction}
Doubly heavy baryons offer a unique opportunity to study QCD in the
presence of heavy quarks as well as providing, through their decays,
information on the weak sector of the Standard Model.  From the
experimental point of view the SELEX Collaboration claimed evidence
for the $\Xi^+_{cc}$ baryon, in the
$\Lambda_c^+K^-\pi^+$~\cite{mattson02} and $pD^+K^-$~\cite{ochera05}
decay modes. The combined analysis gave a mass of
$M_{\Xi^+_{cc}}=3518.7 \pm 1.7\ \mathrm{MeV/c^2}$. However, other
experimental collaborations like FOCUS~\cite{focus03}, {\sl BABAR}
~\cite{babar06} and BELLE~\cite{lesiak06} found no evidence for doubly
charmed baryons and the $\Xi^+_{cc}$ has only been assigned a one star
status by the Particle Data Group (PDG)~\cite{pdg10}. Furthermore, no
evidence for the $\Omega_{cc}^+$ has been reported so
far. 
Nevertheless, being the lightest among the doubly heavy baryons,
one expects doubly charmed baryons masses and decay properties to be
measured in the near future.

While there are many different theoretical determinations of the
doubly charmed baryon masses~\cite{Gershtein:1998sx,Kiselev:1999zj,
Itoh:2000um,Matrasulov:2000us,Gershtein:2000nx, Kiselev:2000jb,
lewis01, Kiselev:2002iy,Ebert:2002ig, Mathur:2002ce,
Flynn:2003vz,Vijande:2004at, Ebert:2005ip,Mehen:2006vv,
Albertus:2006wb, Martynenko:2007je,
Zhang:2008rt,Giannuzzi:2009gh,Albertus:2009ww,Liu:2009jc,
Narison:2010py, weng}, that range from non-relativistic quark model
calculations to unquenched lattice QCD, there are just a few studies
of their decays.

 Total decay widths were evaluated in
Refs.~\cite{Kiselev:1998sy,
Guberina:1999mx,Kiselev:2001fw,Chang:2007xa}, and total semileptonic
and non-leptonic decay rates were predicted in
Ref.~\cite{Guberina:1999mx}.  Some exclusive non-leptonic as well as
semileptonic decay rates of the $\Xi_{cc}$ baryon were calculated in
~\cite{Kiselev:2001fw}. Finally the decay $\Xi_{cc}\to \Xi'_c
e^+\nu_e$ was analyzed in Ref.~\cite{Faessler:2001mr} \footnote{Note
that the $\Xi'_c$ baryon here is denoted as $\Xi_c$ in
Ref.~\cite{Faessler:2001mr}.}. To our knowledge, there is not exist any
systematic study of the exclusive semileptonic $c\to s$ and $c\to d$ 
decay channels of the $\Xi_{cc}$ and $\Omega_{cc}$
baryons. This is the purpose of this work, where we shall
concentrate in transitions to the lowest-lying, $1/2^+$
or $3/2^+$, single-$c$ baryons in the final state. Besides, we will
pay a special attention to possible violations of heavy quark spin
symmetry relations among the relevant form factors, which one might
expect to be sizable at the charm mass scale.

In Table~\ref{tab:baryons}, we show the quantum numbers of the baryons
involved in our calculation. Quark model masses have been taken from
our previous works in Refs.~\cite{Albertus:2003sx,Albertus:2009ww},
where they were obtained using the AL1 potential of
Refs.~\cite{semay94,silvestre96}. Experimental masses are the ones
quoted by the PDG and in the table we quote the average over the
different charge states.  With the exception of the $\Xi_{cc}$, the
agreement is fairly good. For the actual calculation of the decays we
shall use experimental masses except for the $\Xi_{cc}$, which is not well
established, and for the $\Omega_{cc}$ due to the absence 
 of experimental data. In those two cases, we take our model predictions 
 in Table~\ref{tab:baryons} which are in
agreement with different lattice
estimates~\cite{lewis01,Flynn:2003vz,Liu:2009jc}.

The paper is organized as follows: In Sec.~\ref{sect:dwff} we give
general formulae for the semileptonic decay width and the form factor
decomposition of the hadronic matrix elements of the weak current. 
In Sec.~\ref{sect:hqss} 
we will find out heavy quark spin symmetry relations between different 
form factors. Finally in Sec.~\ref{sect:rd} we
present the results. The paper contains also two appendices: In
Appendix~\ref{app:nrbs} we give  a brief description of the
baryon states within the model and  the expressions for the wave functions
of the different baryons and in Appendix~\ref{app:ffwme} we relate the
form factors to weak matrix elements and show how the latter ones are
evaluated in the model.
\begin{table}
\begin{tabular}{ccccccc}\hline\hline
Baryon &~~~~$J^P$~~~~&~~~~ $I$~~~~&~~~~$S^\pi$~~~~& 
Quark content &\multicolumn{2}{c}{Mass\ [MeV]}\\\cline{6-7}

       &       &         &   &      & Quark model   & Experiment                
\\       &       &         &   &      & 
\cite{Albertus:2003sx,Albertus:2009ww}   & \cite{pdg10} \\  
\hline
$\Xi_{cc}$ &$\frac12^+$& $\frac12$ &$1^+$&$ccn$&3613&3518.9
\\
$\Omega_{cc}$  &$\frac12^+$& 0 &$1^+$&$ccs$&3712&--\\\hline
$\Lambda_c$ &$\frac12^+$& 0 &$0^+$&$udc$&2295&2286.5
\\
$\Sigma_c$ &$\frac12^+$& 1 &$1^+$&$nnc$&2469&2453.6
\\
$\Sigma^*_c$ &$\frac32^+$& 1 &$1^+$&$nnc$&2548&2518.0
\\
$\Xi_c$  &$\frac12^+$&$\frac12$&$0^+$&$nsc$&2474&2469.3
\\
$\Xi'_c$  &$\frac12^+$&$\frac12$&$1^+$&$nsc$&2578&2576.8
\\
$\Xi^*_c$ &$\frac32^+$&$\frac12$&$1^+$&$nsc$&2655&2645.9
\\
$\Omega_c$  &$\frac12^+$& 0 &$1^+$&$ssc$&2681&2695.2
\\
$\Omega^*_c$  &$\frac32^+$& 0 &$1^+$&$ssc$&2755&2765.9
\\\hline
\hline
\end{tabular}
\caption{Quantum numbers of double-$c$ and single-$c$ heavy baryons
 involved in this study.  $J^\pi$ and $I$ are the spin-parity and
 isospin of the baryon, while $S^\pi$ is the spin-parity of the two
 heavy or the two light quark subsystem. $n$ denotes a $u$ or $d$
 quark. }
\label{tab:baryons}
\end{table}

\section{Decay width and form factor decomposition of the hadronic current}
\label{sect:dwff}
The total decay width for semileptonic $c\to l$ transitions, with
$l=s,d$, is given by \bea \Gamma&=&|V_{cl}|^2
\frac{G_F^{\,2}}{8\pi^4}\frac{M'^2}{M} \int\sqrt{w^2-1}\, {\cal
L}^{\alpha\beta}(q) {\cal H}_{\alpha\beta}(P,P')\,dw \eea where
$|V_{cl}|$ is the modulus of the corresponding
Cabibbo--Kobayashi--Maskawa (CKM) matrix element for a $c\to l$ quark
transition, for which we shall use $|V_{cs}|=0.97345$ and
$|V_{cd}|=0.2252$ taken from Ref.~\cite{pdg10}.  $G_F= 1.16637(1)\times
10^{-11}$\,MeV$^{-2}$~\cite{pdg10} is the
Fermi decay constant, $P,M$ ($P',M'$) are the four-momentum and mass of
the initial (final) baryon, $q=P-P'$ and $w$ is the product of the
baryons four-velocities
$w=v\cdot v'=\frac{P}M \cdot \frac{P'}{M'}=\frac{M^2+M'^2-q^2}{2MM'}$. In the
decay, $w$ ranges from $w=1$, corresponding to zero recoil of the
final baryon, to a maximum value given, neglecting the neutrino mass,
by $w=w_{\rm max}= \frac{M^2 + M'^2-m^2}{2MM'}$, which depends on the
transition and where $m$ is the final charged lepton mass. Finally
${\cal L}^{\alpha\beta}(q)$ is the leptonic tensor after integrating
in the lepton momenta and ${\cal H}_{\alpha\beta}(P,P')$ is the
hadronic tensor.

The leptonic tensor is given by
\bea
{\cal L}^{\alpha\beta}(q)=A(q^2)\,g^{\alpha\beta}+
B(q^2)\,\frac{q^\alpha q^\beta}{q^2}
\eea
where
\bea
A(q^2)=-\frac{I(q^2)}{6}\left(2q^2-m^2-\frac{m^4}{q^2}\right)\ ,\ \ 
B(q^2)=\frac{I(q^2)}{3}\left({q^2+m^2}-2\frac{m^4}{q^2}\right)
\end{eqnarray}
 with
\begin{eqnarray}
I(q^2)=\frac{\pi}{2q^2}(q^2-m^2)
\end{eqnarray}

The hadronic tensor reads
\begin{eqnarray}
{\cal H}^{\alpha\beta}(P,P') &=& \frac{1}{2J+1} \sum_{r,r'}  
 \big\langle B', r'\
\vec{P}^{\,\prime}\big| J_{cl}^\alpha(0)\big| B, r\ \vec{P}   \big\rangle 
\ \big\langle B', r'\ 
\vec{P}^{\,\prime}\big|J_{cl}^\beta(0) \big|  B, r\ \vec{P} \big\rangle^*
\label{eq:wmunu}
\end{eqnarray}
with $J$ the initial baryon spin, $\big|B, r\ \vec P\big\rangle\,
\left(\big|B', r'\ \vec{P}\,'\big\rangle\right)$ the initial (final)
baryon state with three-momentum $\vec P$ ($\vec{P}\,'$) and spin
third component $r$ ($r'$) in its center of mass frame. $J_{cl}^\mu(0)$ is
the charged weak current for a $c\to l$ quark transition \be
J_{cl}^\mu(0)=\bar\Psi_{l}(0)\gamma^\mu(1-\gamma_5)\Psi_c(0) \ee
Baryonic states are normalized such that \bea \big\langle B, r'\
\vec{P}'\, |\,B, r \ \vec{P} \big\rangle = 2E\,(2\pi)^3
\,\delta_{rr'}\, \delta^3 (\vec{P}-\vec{P}^{\,\prime}) \eea with $E$
the baryon energy for three-momentum $\vec P$.
\subsection{Form factors for  $1/2\to 1/2$ and $1/2\to 3/2$ transitions}
Hadronic matrix elements  can be parameterized in terms of form factors.
For $1/2 \to 1/2$ transitions the commonly used form factor decomposition reads
\begin{eqnarray}
\label{eq:1212}
\big\langle B'(1/2), r'\ \vec{P}^{\,\prime}\left|\,
\overline \Psi_l(0)\gamma^\mu(1-\gamma_5)\Psi_c(0)
 \right| B(1/2), r\ \vec{P}
\big\rangle& =& {\bar u}^{B'}_{r'}(\vec{P}^{\,\prime})\Big\{
\gamma^\mu\left[F_1(w)-\gamma_5 G_1(w)\right]+ v^\mu\left[F_2(w)-\gamma_5
G_2(w)\right]\nonumber\\
&&\hspace{1.5cm}+v'^\mu\left[F_3(w)-\gamma_5 G_3(w)
\right]\Big\}u^{B}_r(\vec{P}\,) \label{eq:def_ff}
\end{eqnarray}
 The $u_{r}$ are Dirac spinors normalized as $({ u}_{r'})^\dagger u_r
 = 2E\,\delta_{r r'}$. $v^\mu$, $v'^\mu $ are the four velocities of
 the initial and final baryons. The three vector $F_1,\,F_2,\,F_3$ and
 three axial $G_1,\,G_2,\,G_3$ form factors are functions of $w$
 or equivalently of $q^2$.\\

For $1/2 \to 3/2$ transitions we follow Llewellyn
 Smith~\cite{Llewellyn Smith:1971zm} to write
\begin{eqnarray}
\label{eq:1232}
&&\hspace{-1cm}\big\langle B'(3/2),r'\vec P'\,|\,\overline 
\Psi_l(0)\gamma^\mu(1-\gamma_5)\Psi_c(0)\,|\,B(1/2),r\,
\vec P\,\big\rangle=
~\bar{u}^{B'}_{\lambda\,r'}(\vec{P}\,')\,\Gamma^{\lambda\mu}(P,P')\,
u^{B}_r(\vec{P}\,)
\nonumber\\
\Gamma^{\lambda\mu}(P,P')=&&
\left[\frac{C_3^V(w)}{M}(g^{\lambda\,\mu}q
\hspace{-.15cm}/\,
-q^\lambda\gamma^\mu)+\frac{C_4^V(w)}{M^2}(g^{\lambda\,\mu}qP'-q^\lambda
P'^\mu)+\frac{C_5^V(w)}{M^2}(g^{\lambda\,\mu}qP-q^\lambda
P^\mu)+C_6^V(w)g^{\lambda\,\mu}\right]\gamma_5\nonumber\\
&&+\left[\frac{C_3^A(w)}{M}(g^{\lambda\,\mu}q
\hspace{-.15cm}/\,
-q^\lambda\gamma^\mu)+\frac{C_4^A(w)}{M^2}(g^{\lambda\,\mu}qP'-q^\lambda
P'^\mu)+{C_5^A(w)}g^{\lambda\,\mu}+\frac{C_6^A(w)}{M^2}
q^\lambda q^\mu\right]
\end{eqnarray}
Here $u^{B'}_{\lambda\,r'}$ is the Rarita-Schwinger spinor of the final spin
3/2 baryon normalized such that $(u_{\lambda\,r'}^{B'})^{\dagger}
u^{B'\,\lambda}_r = -2E'\,\delta_{rr'}$, and we have four vector
($C^V_{3,4,5,6}(w)$) and four axial ($C^A_{3,4,5,6}(w)$) form
factors.\\

In appendix~\ref{app:ffwme} we give the expressions that relate the
form factors to weak current matrix elements and show how the latter
ones are evaluated within the model.

\section{Heavy quark spin symmetry}
\label{sect:hqss}
In hadrons with a single heavy quark the dynamics of the light degrees
of freedom becomes independent of the heavy quark flavour and spin when
the mass of the heavy quark is much larger than $\Lambda_{QCD}$ and
the masses and momenta of the light quarks. This is the essence of
heavy quark symmetry (HQS)~\cite{hqs1,hqs2,hqs3,hqs4}. However, HQS
can not be directly applied to hadrons containing two heavy
quarks. The static theory for a system with two heavy quarks has
infra-red divergences which can be regulated by the kinetic energy
term $\bar h_Q (D^2/2 m_Q) h_Q$. This term breaks the heavy quark
flavour symmetry, but not the spin symmetry for each heavy quark
flavour~\cite{thacker91}. This is known as heavy quark spin symmetry
(HQSS). HQSS implies that all baryons listed in
Table~\ref{tab:baryons} with the same flavour
wave-function  are degenerate.  The invariance of
the effective Lagrangian under arbitrary spin rotations of the $c$
quark leads to relations, near the zero recoil point ($w=1
\leftrightarrow q^2=(M-M')^2 \leftrightarrow |\vec{q}\,|=0$), between
the form factors for vector and axial-vector currents between the
$\Xi_{cc}$ and $\Omega_{cc}$ baryons and the single charmed baryons
listed in Table~\ref{tab:baryons}. These decays are induced by the
semileptonic weak decay of the $c$ quark to a $d$ or a $s$ quark. The
consequences of spin symmetry for weak matrix elements can be derived
using the ``trace formalism''~\cite{Falk:1990yz,MWbook}. To represent
the lowest-lying $S$-wave $ccl$ baryons we will use wave-functions
comprising tensor products of Dirac matrices and spinors,
namely~\cite{Flynn:2007qt}\footnote{We will give here expressions only
for the $c\to d$ transitions of the $\Xi_{cc}$ baryon. Expressions for
the $\Omega_{cc}$ initial baryon and/or $c\to s$ transitions are
totally similar, and SU(3) flavour symmetry could be used to establish
relations between the former and the latter ones.}:
\begin{align}
\label{eq:Xicc}
\Xi_{cc} &=
 -\sqrt{\frac13}
 \left[\frac{(1+\vsl)}2 \g5\right]_{\alpha\beta} u_\gamma(v,r)
\end{align}
where we have indicated Dirac indices $\alpha$, $\beta$ and $\gamma$
explicitly on the right-hand side and $r$ is a helicity label for the
baryon. Under a Lorentz transformation, $\Lambda$, and a $c$
quark spin transformation $S_c$, this wave-function of the form
$\Gamma_{\alpha\beta}\, u_\gamma$ transforms as:
\begin{equation}
\label{eq:spintransfs}
\Gamma\,u \to S(\Lambda) \Gamma S^{-1}(\Lambda)\; S(\Lambda)u,
\quad
\Gamma\,u \to S_c \Gamma \, S_c u.
\end{equation}
The state in Eq.~(\ref{eq:Xicc}) is normalized\footnote{Note, there
are two ways to contract the charm quark indices, leading to $\bar u u
\mathrm{Tr}(\Gamma \overline\Gamma) + \bar u\, \Gamma\,
\overline\Gamma u$, with $\overline\Gamma = \gamma^0 \Gamma^\dagger
\gamma^0$. } to $(-\bar u u=-2M)$, with $M$ the mass of the
state.  On the other hand, the $\Lambda_c$, $\Sigma_c$ and $\Sigma^*_c$
final baryons are represented by the following spinor wave
functions~\cite{MWbook}
\begin{align}
\label{eq:Lambda_c}
\Lambda_{c} &= u_\gamma (v',r') \\
\label{eq:Sigma_c}
\Sigma_{c} &= \left[\frac{1}{\sqrt3} (v^{\prime \lambda} + \gamma^\lambda)
  \g5  u(v',r')\right]_\gamma \\
\label{eq:Sigmastar_c}
\Sigma^*_{c} &= u^\lambda_\gamma (v',r')
\end{align}
For the $\Sigma^*_c$, $u^\lambda_\gamma(v',r')$ is a Rarita-Schwinger
spinor. For $\Sigma_c$, we have taken into account that the light
quarks are coupled to total spin 1 that gives a total spin 1/2 for the
baryon when the spin of the light subsystem is summed with the spin of
the charm quark.  Under a Lorentz transformation , $\Lambda$, and a
$c$ quark spin transformation $S_c$, the above spinor wave functions
transform like $S(\Lambda)\, {\cal U}$ and $S_c \,{\cal U}$, respectively,
with ${\cal U}$ ($= u, \frac{1}{\sqrt3} (v^{\prime \lambda} +
\gamma^\lambda) \g5 u , u^\lambda $) each of the spinors appearing in
Eqs.~(\ref{eq:Lambda_c})--(\ref{eq:Sigmastar_c}). States are
normalized to $\bar u u=2M'$, $(-\bar u u=-2M')$ and $\bar u_\lambda
u^\lambda=-2M'$ for the $\Lambda_c$, $\Sigma_c$ and $\Sigma^*_c$,
respectively.

We can now construct amplitudes for semileptonic $\Xi_{cc} \to
\Lambda_c, \Sigma_c, \Sigma^*_c$ decays, determined by matrix elements
of the weak current $J^\mu = \bar d \gamma^\mu(1-\g5) c$. To that end,
we write the most general form for the matrix element respecting the
heavy quark spin symmetry, taking into account that under a $c$ quark
spin transformation $J^\mu \to J^\mu S_c^\dagger$. We should
distinguish two situations depending on whether  the total spin of the two
light quarks in the final baryon is $S=0$ or $S=1$. In the first (second) case,
the spinor wave--function ${\cal U}$ that represents the final baryon does not
have (has) a Lorentz index. With all these considerations, we have
\begin{eqnarray}
\langle \Lambda_c,v',r'|J^\mu(0)|\Xi_{cc},v,r\rangle &=&\bar
u_{\Lambda_c}(v',r') \frac{(1+\vsl)}2 \g5 \Omega
\gamma^\mu(1-\g5)u_{\Xi_{cc}}(v,r)\\\nonumber &+& \bar
u_{\Lambda_c}(v',r')u_{\Xi_{cc}}(v,r) \mathrm{Tr}[\frac{(1+\vsl)}2 \g5
\Omega \gamma^\mu(1-\g5)] \\ 
\langle\Sigma_c,v',r'|J^\mu(0)|\Xi_{cc},v,r\rangle &=& \bar
u_{\Sigma_c}(v',r')
\frac{1}{\sqrt3}(\gamma^\lambda-v^{\prime\lambda})\g5 \frac{(1+\vsl)}2
\g5 \Omega_\lambda \gamma^\mu(1-\g5)u_{\Xi_{cc}}(v,r)\\\nonumber &+&
\bar
u_{\Sigma_c}(v',r')\frac{1}{\sqrt3}(\gamma^\lambda-v^{\prime\lambda})\g5
u_{\Xi_{cc}}(v,r) \mathrm{Tr}[\frac{(1+\vsl)}2 \g5 \Omega_\lambda
\gamma^\mu(1-\g5)] \\
\langle\Sigma^*_c,v',r'|J^\mu(0)|\Xi_{cc},v,r\rangle &=& \bar
u^\lambda_{\Sigma^*_c}(v',r')
 \frac{(1+\vsl)}2 \g5 \Omega_\lambda
 \gamma^\mu(1-\g5)u_{\Xi_{cc}}(v,r)
\\\nonumber &+&
\bar
u^\lambda_{\Sigma^*_c}(v',r')
u_{\Xi_{cc}}(v,r) \mathrm{Tr}[\frac{(1+\vsl)}2 \g5 \Omega_\lambda
\gamma^\mu(1-\g5)] 
\end{eqnarray}
with\footnote{Terms with a factor of $\vsl$ can be omitted because 
$\vsl (1\pm \vsl)= \pm (1 \pm \vsl)$.}
\begin{eqnarray}
\Omega&=& \beta_1(w) + \beta_2(w)
   \vsl' \\
\Omega_\lambda&=& \delta_1(w) v_\lambda  
+ \delta_2(w)\gamma_\lambda + \delta_3(w)\vsl'v_\lambda+\delta_4
\vsl'\gamma_\lambda
\end{eqnarray}
Note that near the zero recoil point, where the spin symmetry should
work best, HQSS considerably reduces the number of independent form
factors, and it relates those that correspond to transitions where the
spin of the two light quarks in the final baryon is $S=1$. Indeed, we
find at $w=1$
\begin{itemize}
\item $1/2\to 1/2$ transitions ($\Xi_{cc}\to \Lambda_c, \Xi_c$ and 
  $\Omega_{cc}\to \Xi_c$), where the total spin of the two light
quarks in the final baryon is $S=0$:
\begin{eqnarray}
 F_1+F_2+F_3=3G_1\equiv \eta_0
\label{eq:hqss1}
\end{eqnarray}
In the equal mass transition case one would find that $\eta_0$ is normalized as
$\eta_0(w=1)=\sqrt\frac32$. 
\item  Total spin of the two light quarks in the final
baryon is $S=1$ .
\begin{itemize}
\item[*]
$1/2\to 1/2$ transitions ($\Xi_{cc}\to \Sigma_c, \Xi'_c$ and 
  $\Omega_{cc}\to \Xi'_c, \Omega_c$) .
\begin{eqnarray}
F_1+F_2+F_3=\frac35G_1 \equiv \eta_1
\label{eq:hqss2}
\end{eqnarray}
\item[*] $1/2\to 3/2$ transitions ($\Xi_{cc}\to \Sigma^*_c, \Xi^*_c$ and 
  $\Omega_{cc}\to \Xi^*_c, \Omega^*_c$).
\begin{eqnarray}
\frac{\sqrt3}{2}\bigg(C_3^A\frac{M-M'}{M}+
C_4^A\frac{M'(M-M')}{M^2}+C_5^A\bigg) = \eta_1
\label{eq:hqss3}
\end{eqnarray}
\end{itemize}
In the equal mass transition case one would have that $\eta_1(w=1)=\frac1{\sqrt2}$ when
the two light quarks in the final state are different and
$\eta_1(w=1)=1$ when they are equal ($\Omega_c$ and $\Omega^*_c$).
\end{itemize}
Relations (\ref{eq:hqss1}), (\ref{eq:hqss2}) and (\ref{eq:hqss3}) are
exactly satisfied in the quark model when the heavy quark mass is made
arbitrarily large, and thus the calculation is consistent with HQSS
constraints.

\section{Results and discussion} 
\label{sect:rd}

\begin{figure}[t]
\resizebox{5.5cm}{!}{\includegraphics{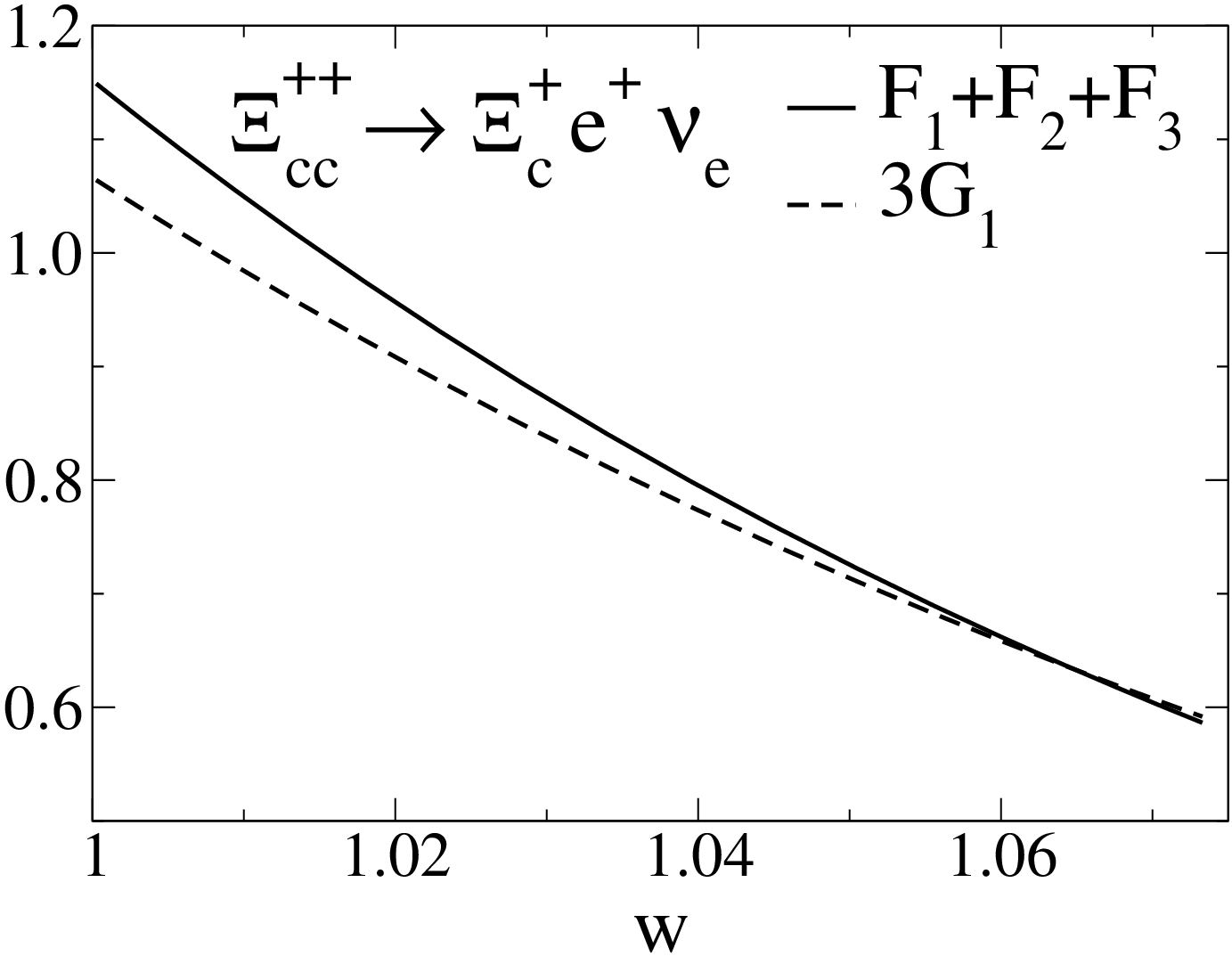}}\hspace{.35cm}
\resizebox{5.5cm}{!}{\includegraphics{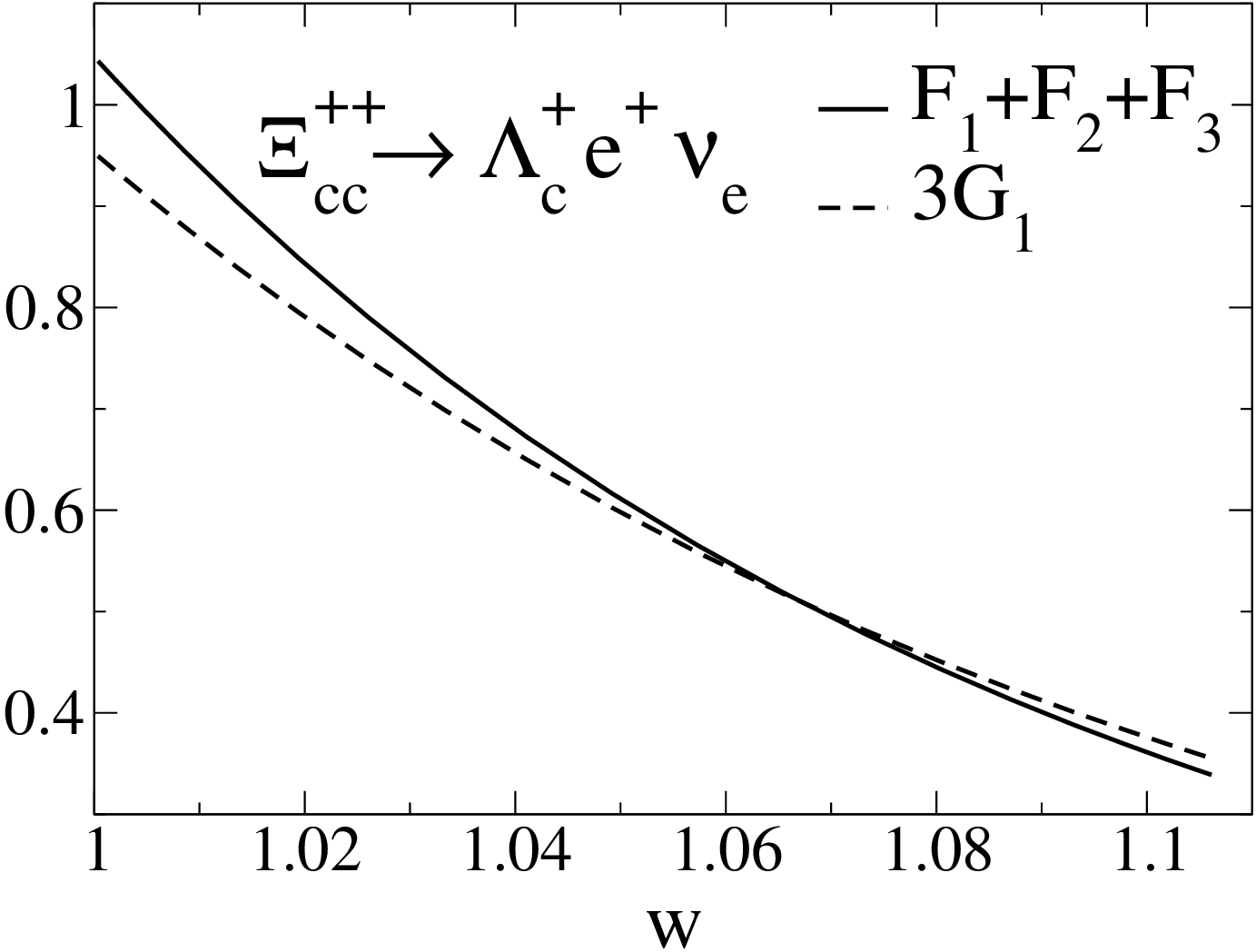}}
\hspace{.25cm}
\resizebox{5.5cm}{!}{\includegraphics{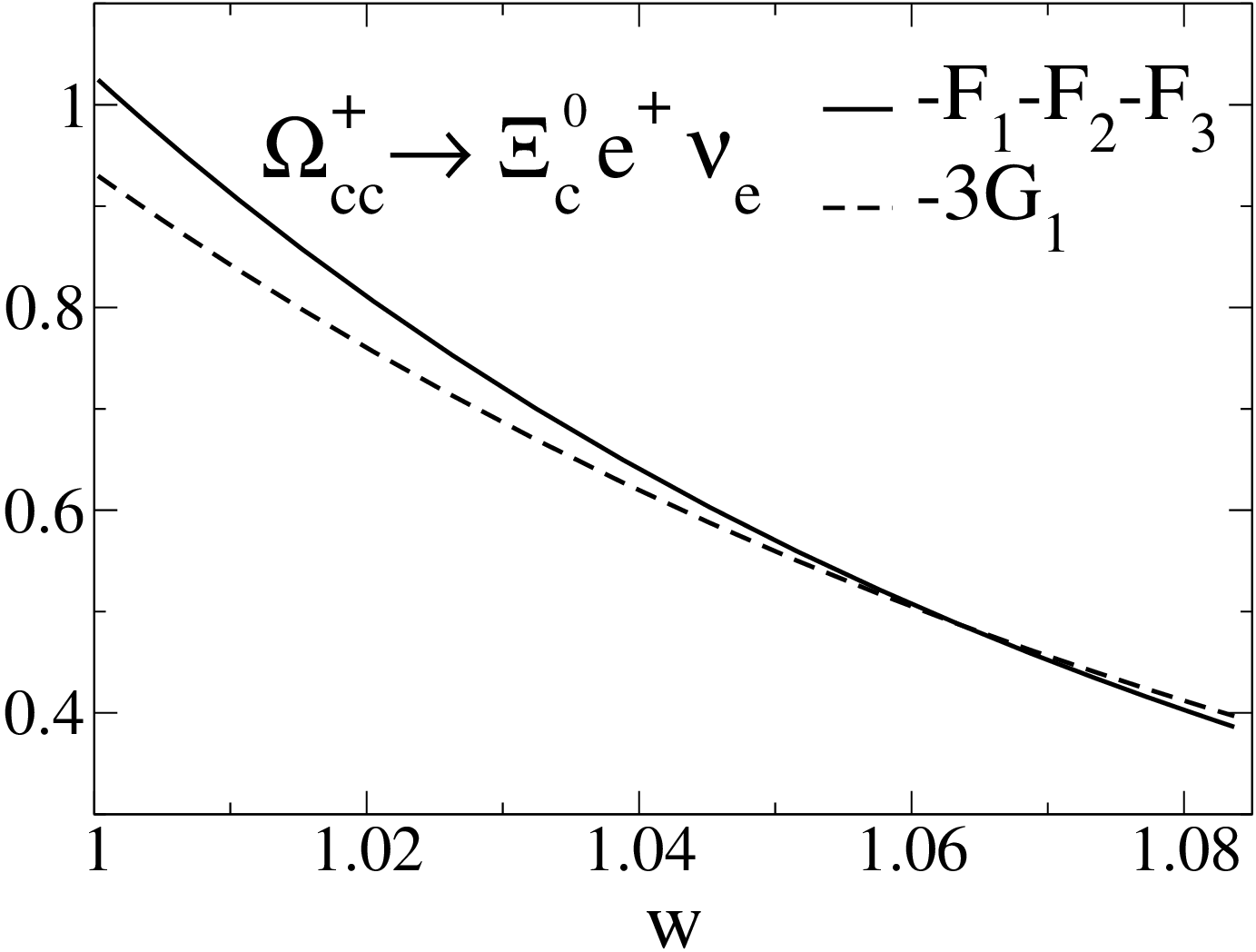}}
\caption{Comparison of $F_1+F_2+F_3$ (solid) and $3G_1$ (dashed) for the
specified transitions. The two light quarks in the final baryon have
total spin $S=0$. In the limit in which the 
heavy quark mass is made arbitrarily large one has that,
near zero recoil ($w=1$), $F_1+F_2+F_3=3G_1$.\vspace{.25cm}
}
\label{fig:s0}
\end{figure}

\begin{figure}[h!!!]
\resizebox{6cm}{!}{\includegraphics{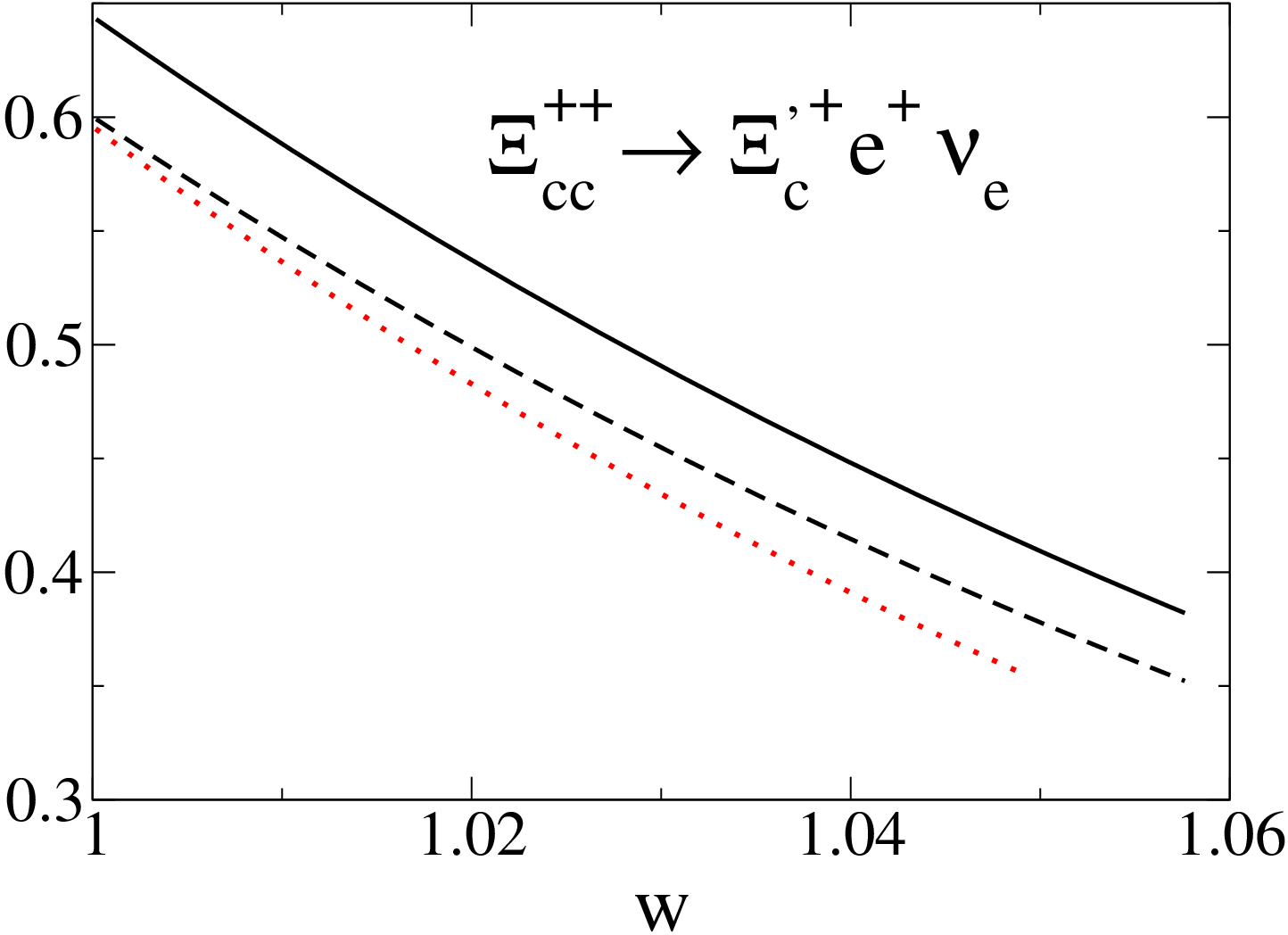}}\hspace{.75cm}
\resizebox{6cm}{!}{\includegraphics{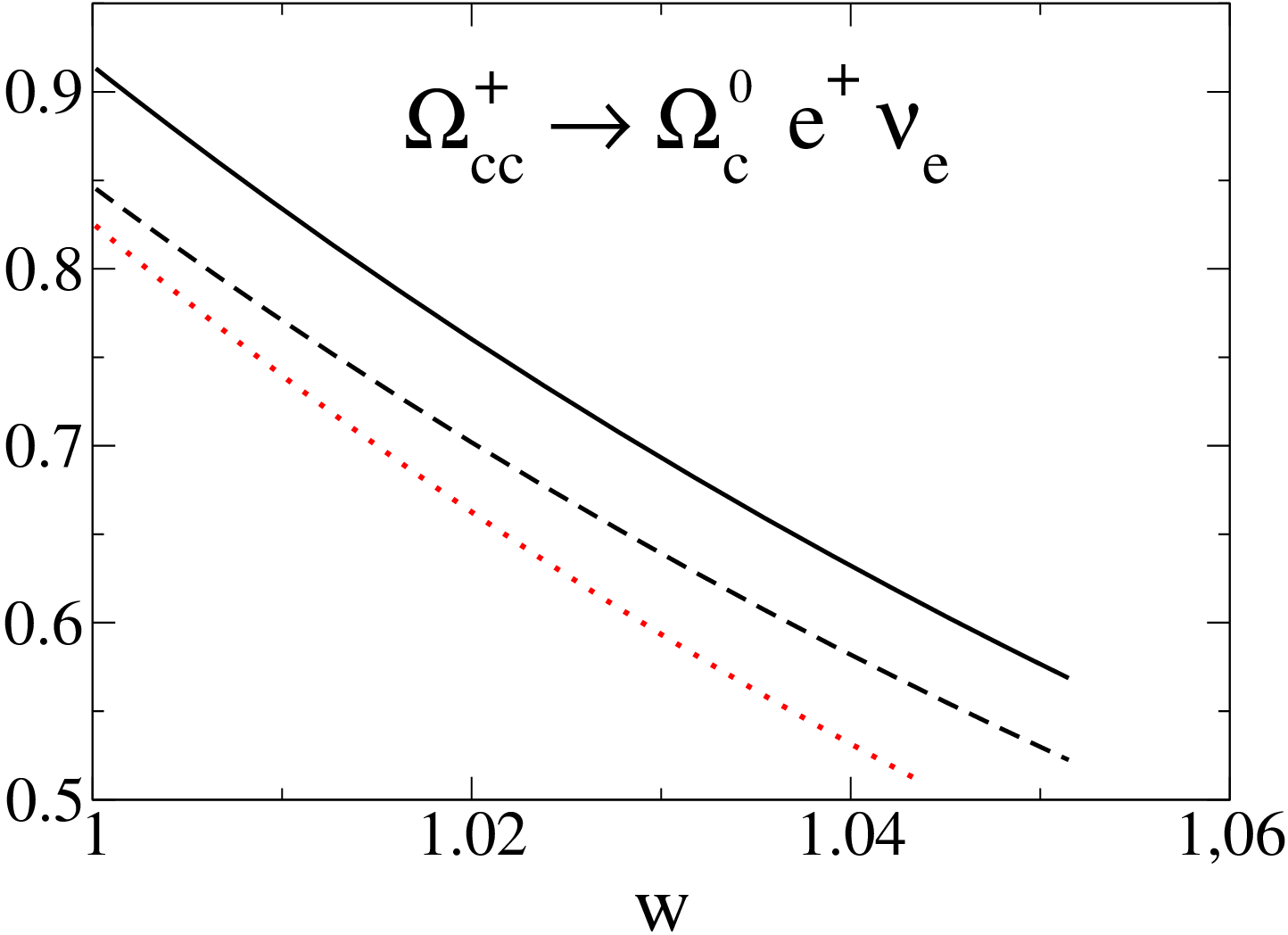}}\vspace{.7cm}\\
\hspace{-.1cm}\resizebox{6.05cm}{!}{\includegraphics{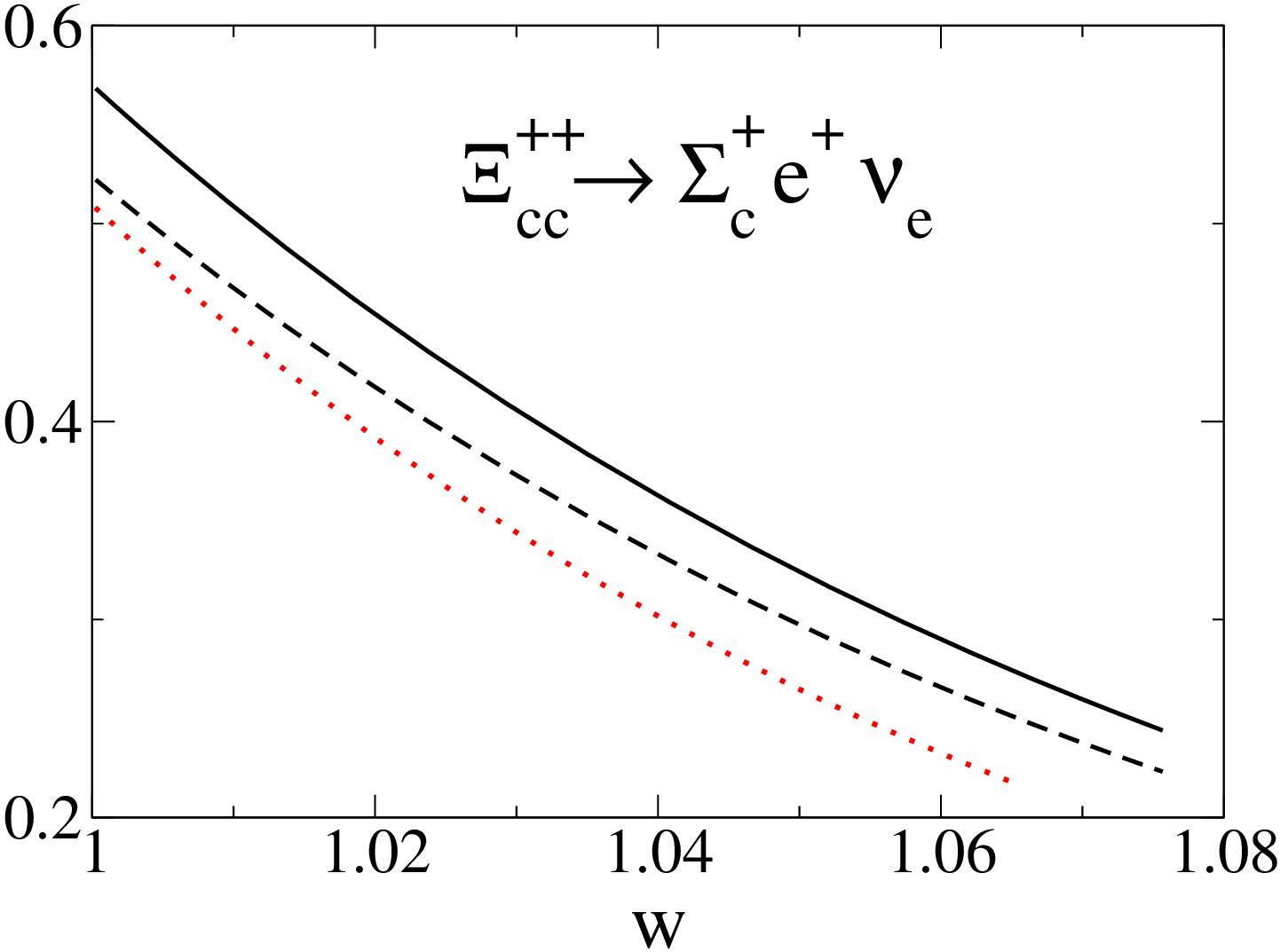}}
\hspace{.75cm}
\resizebox{5.75cm}{!}{\includegraphics{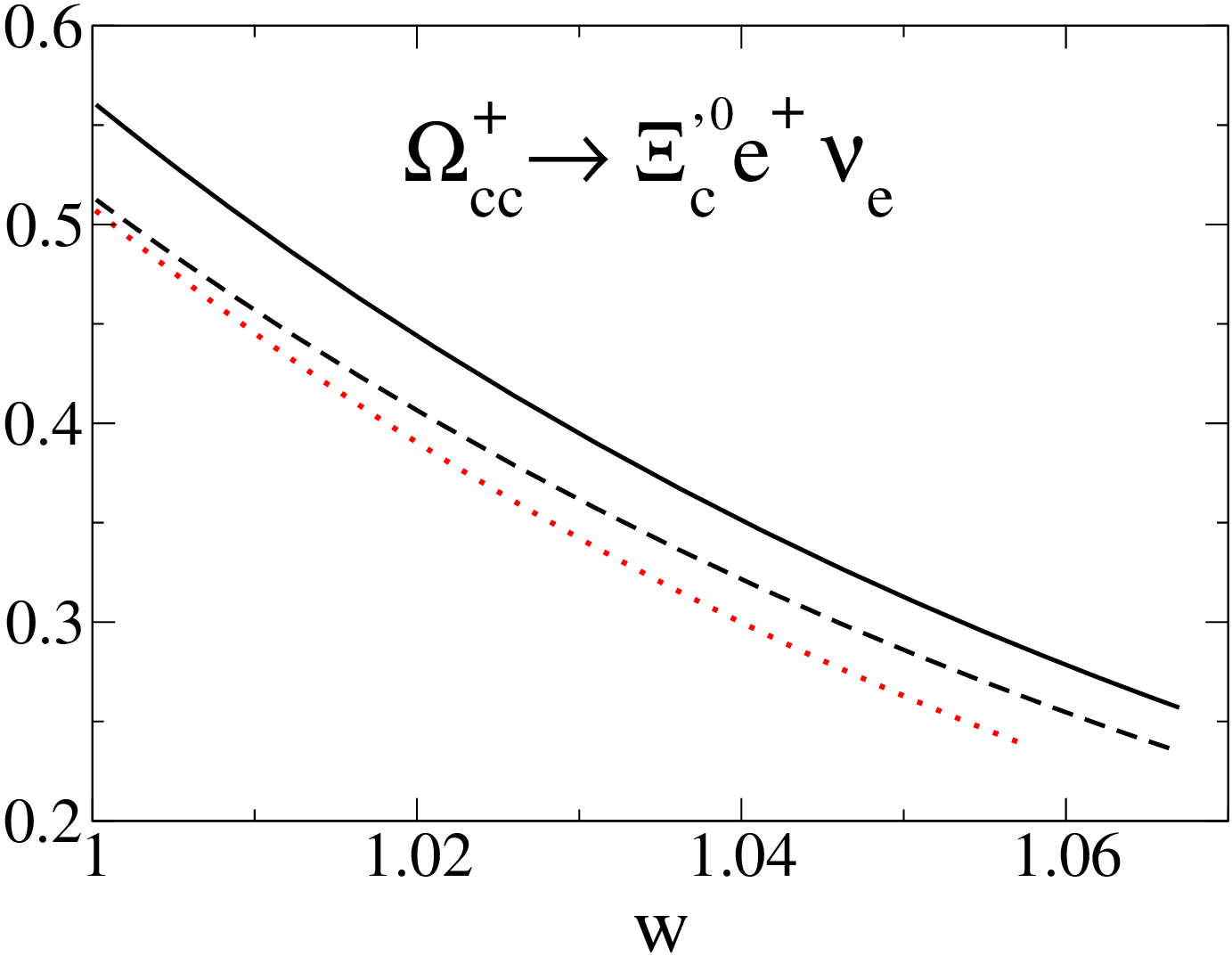}}
\caption{Solid (dashed): $F_1+F_2+F_3$ ($3G_1/5$) for the specified
transitions.  Dotted: the combination
$\frac{\sqrt3}{2}\big(C_3^A\frac{M-M'}{M}
+C_4^A\frac{M'(M-M')}{M^2}+C_5^A\big)$ for the transition with the
corresponding $3/2$ baryon ($\Sigma^*_c$, $\Xi^*_c$ or $\Omega^*_c$)
in the final state. In all cases the two light quarks in the final
baryon have total spin $S=1$. In the limit in which the heavy quark
mass is made arbitrarily large one has that, near zero recoil ($w=1$),
$F_1+F_2+F_3=\frac35G_1= \frac{\sqrt3}{2}\big(C_3^A\frac{M-M'}{M}
+C_4^A\frac{M'(M-M')}{M^2}+C_5^A\big)$.}
\label{fig:s1}
\end{figure}

\begin{figure}[h!!!]
\vspace*{.35cm}\resizebox{6cm}{!}{\includegraphics{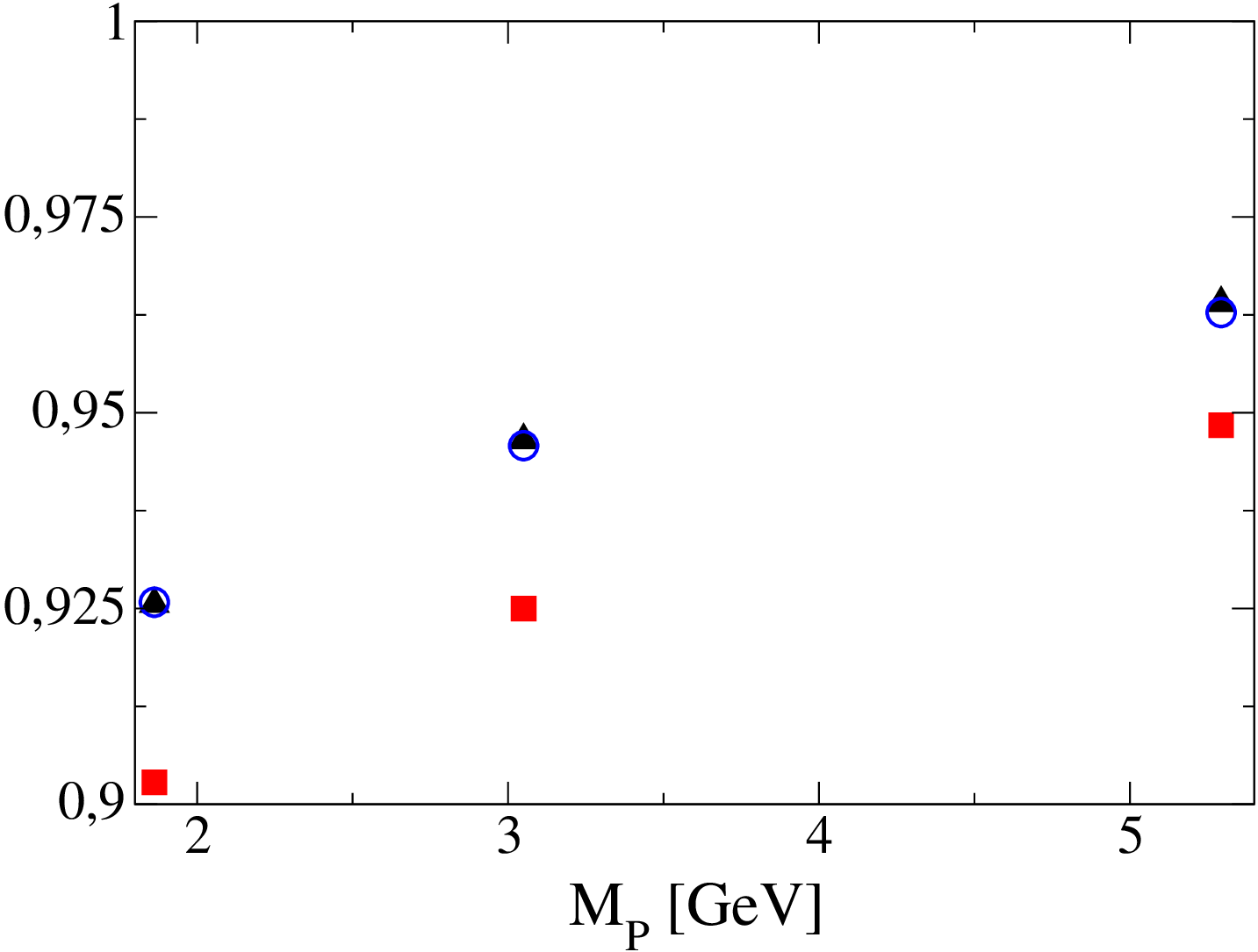}}
\caption{Form factor ratio
$\frac{3G_1}{F_1+F_2+F_3}$ (open circles) from the $\Xi_{cc}^{++}\to\Xi_c^+$ transition, form factor ratio
$\frac{3/5G_1}{F_1+F_2+F_3}$ (up triangles) for the $\Omega_{cc}^{+}\to\Omega_c^{0}$ 
transition and the ratio
$\frac{\frac{\sqrt3}{2}\big(C_3^A\frac{M-M'}{M}
+C_4^A\frac{M'(M-M')}{M^2}+C_5^A\big)}{F_1+F_2+F_3}$ (squares) constructed with the
$C_{3,4,5}^A$ form
factors  from the $\Omega_{cc}^{+}\to\Omega_c^{*0}$ transition and the 
$F_{1,2,3}$ from the $\Omega_{cc}^{+}\to\Omega_c^{0}$ one. Ratios are shown  
as a function of the 
pseudoscalar $P$ heavy-light meson mass for three different heavy quark masses 
and  for $w=1$.}
\label{fig:comparar}
\end{figure}

We start by checking that our calculation respects the constraints on
the form factors deduced from HQSS.
 In Figs.~\ref{fig:s0} and \ref{fig:s1}, we show to what extent the
relations of (\ref{eq:hqss1}), (\ref{eq:hqss2}) and (\ref{eq:hqss3})
deduced above are satisfied for the actual $m_c$ value. In all cases
we see  moderate deviations, that stem from $1/m_c$ corrections, at the
level of  about 10\% near zero recoil, though larger than those
found in \cite{Hernandez:2007qv} for the $b\to c$ transitions of the
$\Xi_{bc}$ and $\Xi_{bb}$ baryons. These discrepancies tend to disappear
when the mass of the heavy quark is made arbitrarily large. This is illustrated in 
Fig.~\ref{fig:comparar} where we show,
for $w=1$ and for three different heavy quark masses, the  form factor ratio
$\frac{3G_1}{F_1+F_2+F_3}$ from the $\Xi_{cc}^{++}\to\Xi_c^+$ transition, the
form factor ratio $\frac{3/5G_1}{F_1+F_2+F_3}$for the $\Omega_{cc}^{+}\to\Omega_c^{0}$ 
transition and the ratio
$\frac{\sqrt3}{2}\big(C_3^A\frac{M-M'}{M}
+C_4^A\frac{M'(M-M')}{M^2}+C_5^A\big)/(F_1+F_2+F_3)$ constructed  with the
$C_{3,4,5}^A$ form
factors  from the $\Omega_{cc}^{+}\to\Omega_c^{*0}$ transition and the 
$F_{1,2,3}$ from the $\Omega_{cc}^{+}\to\Omega_c^{0}$ one. 
The ratios are shown as a
function of the corresponding pseudoscalar $P$ heavy-light meson mass. As the 
pseudoscalar meson  mass
increases (the heavy quark mass increases) the ratios tend to one as expected. Similar results
are obtained in the other cases.
Even though we are not in the infinite heavy quark mass limit, HQSS turns out to be
a useful tool to understand the dynamics of the $c\to s,d$ 
$\Xi_{cc}$ and $\Omega_{cc}$   decays near zero recoil. One also sees
 that at $w=1$ our results for $\eta_0(w=1), \eta_1(w=1)$ are
systematically smaller than would be expected for an equal mass
transition. This is a reflection of the mismatch in the wave functions
due to the different initial ($c$) and final ($d$ or $s$) quark masses
in the $c\to d,s$ decays.

\begin{table}
\begin{tabular}{lcccc}\hline\hline
&&\multicolumn{3}{c}{ $\Gamma\ [\,{\rm ps}^{-1}]$} \vspace{.1cm}\\\cline{3-5}
$B_{cc}\to B_ce^+\nu_e$&\hspace*{.5cm}Quark transition\hspace*{.5cm} 
&This work&\cite{Faessler:2001mr}&\cite{Kiselev:2001fw}\\\hline
$\Xi_{cc}^{++}\to\Xi^+_ce^+\nu_e$&$(c\to s)$&$8.75\times 10^{-2}$\\
$\Xi_{cc}^{+\hspace{.18cm}}\to\Xi^0_ce^+\nu_e$&$(c\to s)$&$8.68\times 10^{-2}$\\
$\Xi_{cc}^{++}\to\Xi'^+_ce^+\nu_e$&$(c\to s)$&0.146&$0.208\div 0.258$\\
$\Xi_{cc}^{+\hspace{.18cm}}\to\Xi'^0_ce^+\nu_e$&$(c\to s)$&0.145&$0.208\div 0.258$\\
$\Xi_{cc}^{++}\to\Xi^{*\,+}_ce^+\nu_e$&$(c\to s)$&$3.20\times 10^{-2}$\\
$\Xi_{cc}^{+\hspace{.18cm}}\to\Xi^{*\,0}_ce^+\nu_e$&$(c\to s)$&$3.20\times 10^{-2}$\\
$\Xi_{cc}^{++}\to\Xi'^+_ce^+\nu_e+\Xi^+_ce^+\nu_e+\Xi^{*\,+}_ce^+\nu_e$&
$(c\to s)$&0.266&&$0.37\pm0.04^{(*)}$\\
$\Xi_{cc}^{+\hspace{.18cm}}\to\Xi'^0_ce^+\nu_e\ +\Xi^0_ce^+\nu_e
\ +\Xi^{*\,0}_ce^+\nu_e$&$(c\to
s)$&0.264&&$0.47\pm0.15^{(*)}$\\
$\Xi_{cc}^{++}\to\Lambda^+_ce^+\nu_e$\hspace*{.5cm}&$(c\to d)$&$4.86\times
10^{-3}$\\
$\Xi_{cc}^{++}\to\Sigma^+_ce^+\nu_e$&$(c\to d)$&$7.94\times 10^{-3}$\\
$\Xi_{cc}^{+\hspace{.18cm} }\to\Sigma^0_ce^+\nu_e$&$(c\to d)$&$1.58\times 10^{-2}$\\
$\Xi_{cc}^{++}\to\Sigma^{*\,+}_ce^+\nu_e$&$(c\to d)$&$1.77\times 10^{-3}$\\
$\Xi_{cc}^{+\hspace{.18cm}}\to\Sigma^{*\,0}_ce^+\nu_e$&$(c\to d)$&$3.54\times
10^{-3}$\\
\hline
$\Omega_{cc}^{+\hspace{.18cm}}\to\Omega^{0}_ce^+\nu_e$&$(c\to s)$&0.282\\
$\Omega_{cc}^{+\hspace{.18cm}}\to\Omega^{*\,0}_ce^+\nu_e$&$(c\to s)$&$5.77\times 10^{-2}$\\
$\Omega_{cc}^{+\hspace{.18cm}}\to\Xi^{0}_ce^+\nu_e$&$(c\to d)$&$4.11\times
10^{-3}$\\
$\Omega_{cc}^{+\hspace{.18cm}}\to\Xi'^{0}_ce^+\nu_e$&$(c\to d)$&$7.44\times
10^{-3}$\\
$\Omega_{cc}^{+\hspace{.18cm}}\to\Xi^{*\,0}_ce^+\nu_e$&$(c\to d)$&$1.72\times
10^{-3}$\\
\hline\hline
\end{tabular}
\caption{Decay widths in units of ${\rm ps}^{-1}$. We use
$|V_{cs}|=0.97345$ and $|V_{cd}|=0.2252$ taken from Ref.~\cite{pdg10}. Results 
with an ${(\ast)}$, our
estimates from the total decay widths and branching ratios
in~\cite{Kiselev:2001fw}. Similar results are obtained for
$\mu^+\nu_\mu$ leptons in the final state.}
\label{tab:dw}
\end{table}

Now we discuss the results for the decay widths. Those are shown in
Table~\ref{tab:dw} for the dominant ($c\to s$) and sub-dominant ($c\to d$)
exclusive semileptonic decays of the $\Xi_{cc}$ and $\Omega_{cc}$ to ground
state, $1/2^+$ or $3/2^+$, single charmed baryons and with a positron
in the final state\footnote{ Similar results are obtained for
$\mu^+\nu_\mu$ leptons in the final state.}.  For the $\Omega^+_{cc}$
baryon, semileptonic decays driven by a $s\to u$ transition at the
quark level are also possible.  However, in this latter case phase
space is very limited and we find the decay widths are orders of
magnitude smaller than the ones shown.  To our knowledge there are
just a few previous theoretical evaluations of the $\Xi_{cc}$ semileptonic
decays. In Ref.~\cite{Faessler:2001mr} the authors use the
relativistic three-quark model to evaluate the $\Xi_{cc}\to \Xi'_c
e^+\nu_e$ decay, while in Ref.~\cite{Kiselev:2001fw}, using heavy
quark effective theory and non-relativistic QCD sum rules, they give
both the lifetime of the $\Xi_{cc}$ baryon and the branching ratio for
the combined decay $\Xi_{cc}\to\Xi_c e^+\nu_e+\Xi'_c
e^+\nu_e+\Xi_c^*e^+\nu_e$ from which we have evaluated the
semileptonic decay widths shown in the table. We find a fair agreement
of our predictions with both calculations. In
Ref.~\cite{Guberina:1999mx}, using the optical theorem and the
operator product expansion, the authors evaluated the total
semileptonic decay rate finding it to be $0.151\,{\rm ps}^{-1}$ for
$\Xi_{cc}^{++}$ and $0.166\,{\rm ps}^{-1}$ for $\Xi_{cc}^{+}$. These
values are roughly a factor of two smaller than the sum of our partial
decay widths or the results in Ref.~\cite{Kiselev:2001fw}.  For the
$\Omega_{cc}^+$ a total semileptonic decay width of $0.454\,{\rm
ps}^{-1}$ is given in Ref.~\cite{Guberina:1999mx}. In this case this
is in better agreement with the sum of our partial semileptonic decay
widths which add up to $0.353\,{\rm ps}^{-1}$. 

An estimate of part of the
uncertainties in our model can be done by evaluating the decay widths using
wave functions produced with different interquark interactions. We have done
this by using the AP1~\cite{semay94,silvestre96} and Bhaduri~\cite{BD81}
interquark potentials finding changes in the decay widths to be at the level of $10\,$\%.
Another source of uncertainties may come from the 
contribution from intermediate heavy-light vector meson ($D^*$ and
$D^*_s$) exchanges~\cite{Isgur:1989qw}. They are neither considered in this
work nor in the previous quark model calculation of
Ref.~\cite{Faessler:2001mr}\footnote{We think, these effects are not
explicitly taken into account either in the QCD sum rule approach of
Ref.~\cite{Kiselev:2001fw} or in that, based in the optical theorem, 
followed in ~\cite{Guberina:1999mx}.}. We expect such exchanges
to produce small effects\footnote{Moreover in the transitions studied
  here, 
the intermediate vector mesons would be far off shell. Thus, the
uncertainties related to the strength of their couplings with the
singly and doubly charmed baryons, and those stemming from the lack of
a reasonable scheme to model how the latter interactions are
suppressed when $q^2$ approaches  the endpoint of the available
phase-space ($q^2=0$) would make meaningless the computation of these
effects.} in the integrated widths, specially for the decays
considered in this work, for which the $D^*$ and $D^*_s$ poles are located far
from $\sqrt{q^2_{\rm max}}$. This is in sharp contrast with the
situation for the $B \to \pi $ and $D \to \pi$
decays~\cite{Isgur:1989qw,Albertus:2005ud}. The model could  be also improved 
by considering two body operators, and going in this manner beyond the 
spectator approximation. However, two body current
contributions are not straightforward to compute, and since we expect
moderate effects\footnote{The
difference between the sum of masses of the constituent quarks and
that of the baryon provides a first estimate of these
effects~\cite{Albertus:2004wj}.},  similar to the other
uncertainties mentioned above, we will leave this issue for future
research. Moreover, there exists a greater source of uncertainties
affecting  our results that comes from our limited knowledge on the
masses of the initial double charmed baryons. As we pointed out in the
introduction, for the $\Xi_{cc}$ and the $\Omega_{cc}$ baryons, we
have used our quark model predictions in Table~\ref{tab:baryons}. If
the SELEX Collaboration measured mass for the $\Xi_{cc}$ baryon is
used instead, we would find significantly smaller decay widths by about 20\%. 
This is just because of the reduction on the
 available  phase-space for the decay. None of the  theoretical works
mentioned in  Table~\ref{tab:dw}  use
the SELEX mass value.

To summarize this work, we would like to point out that we have
carried out the first systematic study of all dominant and sub-dominant
semi-leptonic transitions of the doubly charmed $\Xi_{cc}$ and
$\Omega_{cc}$ baryons to the lowest-lying, $1/2^+$ or $3/2^+$,
single-$c$ baryons. To that end, we have employed a simple constituent
quark model scheme, which benefits from the important
simplifications~\cite{Albertus:2006wb,Albertus:2003sx} of the
non-relativistic three body problem that stem from the application of
HQSS. We have also derived, for the first time, HQSS relations among
the relevant form factors that govern these decays near zero recoil,
and have found the size of the deviations induced by the finite charm
quark mass.

Predictions of this framework have been successfully tested in
the past in the context of the $\Lambda_b$ and $\Xi_b$ semileptonic
decays~\cite{Albertus:2004wj}. There, we obtained results for partially
integrated decay widths that nicely compared with lattice
results~\cite{HB-Lattice}, and  from  
the experimental $\Lambda_b-$semileptonic decay, we could  also
determine the $V_{cb}$ CKM matrix element in
excellent agreement with the accepted values quoted in the PDG~\cite{pdg10}.

\begin{acknowledgments}
  This research was supported by DGI and FEDER funds, under contracts
  FIS2006-03438, FIS2008-01143/FIS, FPA2010-21750-C02-02, and the Spanish
  Consolider-Ingenio 2010 Programme CPAN (CSD2007-00042),  by Generalitat
  Valenciana under contract PROMETEO/20090090 and by the EU
  HadronPhysics2 project, grant agreement no. 227431. C. A. thanks a Juan de 
  la Cierva contract from the
Spanish  Ministerio de Educaci\'on y Ciencia.
\end{acknowledgments}

\appendix
\section{Non-relativistic baryon states and wave functions}
\label{app:nrbs}
Our non-relativistic states 
 are constructed as a superposition of three quark states
\begin{eqnarray}
\label{wf}
&&\hspace{-1cm}\big|{B,r\,\vec{P}}\,\big\rangle_{NR}
=\sqrt{2E}\int d^{\,3}Q_1 \int d^{\,3}Q_2\ \frac1{\sqrt2} \sum_{\alpha_1,\alpha_2,\alpha_3}
\hat{\psi}^{(B,r)}_{\alpha_1\,\alpha_2\,\alpha_3}(\,\vec{Q}_1,\vec{Q}_2\,)
\ \frac{1}{(2\pi)^3\ \sqrt{2E_{f_1}2E_{f_2}
2E_{f_3}}}\nonumber\\ 
&&\hspace{3cm}
\times\big|\ \alpha_1\
\vec{p}_1=\frac{m_{f_1}}{\overline{M}}\vec{P}+\vec{Q}_1\ \big\rangle
\big|\ \alpha_2\ \vec{p}_2=\frac{m_{f_2}}{\overline{M}}\vec{P}+\vec{Q}_2\ \big\rangle
\big|\ \alpha_3\ \vec{p}_3=\frac{m_{f_3}}{\overline{M}}\vec{P}-\vec{Q}_1
-\vec{Q}_2\ \big\rangle
 \end{eqnarray}
The factor $\sqrt{2E}$ is introduced for convenience in order to have
the proper normalization.  $\alpha_j$ represents the spin (s), flavour
(f) and color (c) quantum numbers ( $\alpha\equiv (s,f,c)$\,) of the
j-th quark, and $(E_{f_j},\,\vec{p}_j),\, m_{f_j}$ are its
four-momentum and mass. $\overline{M}$ is given by
$\overline{M}=m_{f_1}+m_{f_2}+m_{f_3}$.  Individual quark states are
normalized such that $\left\langle\ \alpha^{\prime}\ \vec{p}^{\
\prime}\,|\,\alpha\ \vec{p}\, \right\rangle=2E_f\, (2\pi)^3\,
\delta_{\alpha^{\prime}\, \alpha}\,\delta^3( \vec{p}^{\
\prime}-\vec{p}\,)$. $\hat{\psi}^{\,(B,r)}_{\alpha_1\,\alpha_2\,\alpha_3}
(\,\vec{Q}_1,\vec{Q}_2\,)$ is the internal wave function in momentum
space, being $\vec{Q}_1$ ($\vec{Q}_2$) the conjugate momenta to the
relative position $\vec{r}_1$ ($\vec{r}_2$) between quark 1 (2) and the
 third quark.  In the transitions under study an initial $c\,c\,l'$
baryon decays into a final $c\,l\,l'$ one, where $l=d,s$ and
$l'=u,d,s$. We construct the wave functions such that the two $c$
quarks in the initial baryon, or the two light quarks in the final
baryon, are quarks 1 and 2.  Expressions for the different
$\hat{\psi}^{(B,r)}_{\alpha_1\,\alpha_2\,\alpha_3}
(\,\vec{Q}_1,\vec{Q}_2\,)$ are given  below.
These wave functions are normalized as
\begin{equation}
\int d^{\,3}Q_1 \int d^{\,3}Q_2\ \sum_{\alpha_1,\alpha_2,\alpha_3}
\left(\hat{\psi}^{(B,r')}_{\alpha_1\,\alpha_2\,\alpha_3}(\,\vec{Q}_1,\vec{Q}_2\,)\right)^*
\hat{\psi}^{(B,r)}_{\alpha_1\,\alpha_2\,\alpha_3}(\,\vec{Q}_1,\vec{Q}_2\,)
=\delta_{rr'}
\end{equation}
so that we get for our non-relativistic baryon states  
%
${}_{\stackrel{}{NR}}
\big\langle\, {B,r'\,\vec{P}^{\,\prime}}\,|\,{B,r
\,\vec{P}}\,\big\rangle_{NR}
=2E\,(2\pi)^3\,\delta_{rr'}\,\delta^3(\vec{P}^{\,\prime}-\vec{P}\,)$.

The wave functions of the different non-strange
states included in this study are given by
\begin{eqnarray}
\hat{\psi}^{\,(\Xi^{++}_{cc},r)}_{\alpha_1\,\alpha_2\,\alpha_3}(\,\vec{Q}_1,\vec{Q}_2\,)
&=&\frac{\varepsilon_{c_1\,c_2\,c_3}}{\sqrt{3!}}\
\ \widetilde{\phi}^{\,(\Xi^{++}_{cc})}(\,\vec{Q}_1,\vec{Q}_2\,)\ 
\delta_{f_1\,c}\,\delta_{f_2\,c}\, \delta_{f_3\,u}\nonumber\\
&&\hspace{2cm} \times\  (1/2,1/2,1;s_1,s_2,s_1+s_2)\ 
(1,1/2,1/2;s_1+s_2,s_3,r)\\
\hat{\psi}^{\,(\Xi^+_{cc},r)}_{\alpha_1\,\alpha_2\,\alpha_3}(\,\vec{Q}_1,\vec{Q}_2\,)
&=&\frac{\varepsilon_{c_1\,c_2\,c_3}}{\sqrt{3!}}\
\ \widetilde{\phi}^{\,(\Xi^+_{cc})}(\,\vec{Q}_1,\vec{Q}_2\,)\ 
\delta_{f_1\,c}\,\delta_{f_2\,c}\, \delta_{f_3\,d}\nonumber\\
&&\hspace{2cm} \times\  (1/2,1/2,1;s_1,s_2,s_1+s_2)\ 
(1,1/2,1/2;s_1+s_2,s_3,r)\\
\hat{\psi}^{\,(\Lambda^+_{c},r)}_{\alpha_1\,\alpha_2\,\alpha_3}(\,\vec{Q}_1,\vec{Q}_2\,)
&=&\frac{\varepsilon_{c_1\,c_2\,c_3}}{\sqrt{3!}}\
\ \widetilde{\phi}^{\,(\Lambda^+_{c})}(\,\vec{Q}_1,\vec{Q}_2\,)\ 
\frac{1}{\sqrt2}(\delta_{f_1\,u}\, \delta_{f_2\,d}-\delta_{f_1\,d}\, 
\delta_{f_2\,u})\, \delta_{f_3\,c}\  (1/2,1/2,0;s_1,s_2,0)\,\delta_{s_3\,r}\\
\hat{\psi}^{\,(\Sigma^{+}_{c},r)}_{\alpha_1\,\alpha_2\,\alpha_3}(\,\vec{Q}_1,\vec{Q}_2\,)
&=&\frac{\varepsilon_{c_1\,c_2\,c_3}}{\sqrt{3!}}\
\ \widetilde{\phi}^{\,(\Sigma^{+}_{c})}(\,\vec{Q}_1,\vec{Q}_2\,)\ 
\frac{1}{\sqrt2}(\delta_{f_1\,u}\, \delta_{f_2\,d}+\delta_{f_1\,d}\, \delta_{f_2\,u})\, \delta_{f_3\,c}\nonumber\\
&&\hspace{2cm} \times\   (1/2,1/2,1;s_1,s_2,s_1+s_2)\ 
(1,1/2,1/2;s_1+s_2,s_3,r)
\eea
\bea
\hat{\psi}^{\,(\Sigma^{0}_{c},r)}_{\alpha_1\,\alpha_2\,\alpha_3}(\,\vec{Q}_1,\vec{Q}_2\,)
&=&\frac{\varepsilon_{c_1\,c_2\,c_3}}{\sqrt{3!}}\
\ \widetilde{\phi}^{\,(\Sigma^{0}_{c})}(\,\vec{Q}_1,\vec{Q}_2\,)\ 
\delta_{f_1\,d}\, \delta_{f_2\,d}\, \delta_{f_3\,c}\nonumber\\
&&\hspace{2cm} \times\   (1/2,1/2,1;s_1,s_2,s_1+s_2)\ 
(1,1/2,1/2;s_1+s_2,s_3,r)\\
\hat{\psi}^{\,(\Sigma^{*\,+}_{c},r)}_{\alpha_1\,\alpha_2\,\alpha_3}(\,\vec{Q}_1,\vec{Q}_2\,)
&=&\frac{\varepsilon_{c_1\,c_2\,c_3}}{\sqrt{3!}}\
\ \widetilde{\phi}^{\,(\Sigma^{*\,+}_{c})}(\,\vec{Q}_1,\vec{Q}_2\,)\ 
\frac{1}{\sqrt2}(\delta_{f_1\,u}\, \delta_{f_2\,d}+\delta_{f_1\,d}\, \delta_{f_2\,u})\, \delta_{f_3\,c}\nonumber\\
&&\hspace{2cm} \times\   (1/2,1/2,1;s_1,s_2,s_1+s_2)\ 
(1,1/2,3/2;s_1+s_2,s_3,r) \\
\hat{\psi}^{\,(\Sigma^{*\,0}_{c},r)}_{\alpha_1\,\alpha_2\,\alpha_3}(\,\vec{Q}_1,\vec{Q}_2\,)
&=&\frac{\varepsilon_{c_1\,c_2\,c_3}}{\sqrt{3!}}\
\ \widetilde{\phi}^{\,(\Sigma^{*\,0}_{c})}(\,\vec{Q}_1,\vec{Q}_2\,)\ 
\delta_{f_1\,d}\, \delta_{f_2\,d}\, \delta_{f_3\,c}\nonumber\\
&&\hspace{2cm} \times\   (1/2,1/2,1;s_1,s_2,s_1+s_2)\ 
(1,1/2,3/2;s_1+s_2,s_3,r) 
\eea
where
$\varepsilon_{c_1 c_2 c_3}$ is the totally antisymmetric tensor
with  $\frac{\varepsilon_{c_1 c_2 c_3}}{\sqrt{3!}}$ being the fully
antisymmetric color wave
function. The $(j_1,j_2,j;m_1,m_2,m)$ are SU(2) Clebsch-Gordan 
coefficients.
The different
   $\tilde \phi(\,\vec{Q}_1,\vec{Q}_2\,)$ wave functions verify
$\tilde \phi(\,\vec{Q}_2,\vec{Q}_1\,)=\tilde \phi(\,\vec{Q}_1,\vec{Q}_2\,)$ and
they have  total orbital angular momentum  0 being invariant under
rotations and thus depending only  on
  $|\vec{Q}_1|$, $|\vec{Q}_2|$ 
and $\vec{Q}_1\cdot\vec{Q}_2$. They are normalized such that
\begin{equation}
\int d^{\,3}Q_1 \int d^{\,3}Q_2\ 
\left|\widetilde{\phi}(\,\vec{Q}_1,\vec{Q}_2\,)\right|^2
=1
\end{equation}
For states with $s$-quark content we further have  
\begin{eqnarray}
\hat{\psi}^{\,(\Omega^{+}_{cc},r)}_{\alpha_1\,\alpha_2\,\alpha_3}(\,\vec{Q}_1,\vec{Q}_2\,)
&=&\frac{\varepsilon_{c_1\,c_2\,c_3}}{\sqrt{3!}}
\ \widetilde{\phi}^{\,(\Omega^{+}_{cc})}(\,\vec{Q}_1,\vec{Q}_2\,)\ 
\delta_{f_1\,c}\,\delta_{f_2\,c}\, \delta_{f_3\,s}\nonumber\\
&&\hspace{2cm} \times\  (1/2,1/2,1;s_1,s_2,s_1+s_2)\ 
(1,1/2,1/2;s_1+s_2,s_3,r)\\
\hat{\psi}^{\,(\Xi^+_{c},r)}_{\alpha_1\,\alpha_2\,\alpha_3}(\,\vec{Q}_1,\vec{Q}_2\,)
&=&\frac{\varepsilon_{c_1\,c_2\,c_3}}{\sqrt{3!}}
\ \frac1{\sqrt2}\,[\widetilde{\phi}^{\,(\Xi^+_{c})}_{us}(\,\vec{Q}_1,\vec{Q}_2\,)\ 
\delta_{f_1\,u}\, \delta_{f_2\,s}
-\widetilde{\phi}^{\,(\Xi^+_{c})}_{su}(\,\vec{Q}_1,\vec{Q}_2\,)\ 
\delta_{f_1\,s}\, \delta_{f_2\,u}
]\, \delta_{f_3\,c}\nonumber\\
&&\hspace{2cm} \times\   (1/2,1/2,0;s_1,s_2,0)\,\delta_{s_3\,r} \\
\hat{\psi}^{\,(\Xi^0_{c},r)}_{\alpha_1\,\alpha_2\,\alpha_3}(\,\vec{Q}_1,\vec{Q}_2\,)
&=&\frac{\varepsilon_{c_1\,c_2\,c_3}}{\sqrt{3!}}
\ \frac1{\sqrt2}\,[\widetilde{\phi}^{\,(\Xi^0_{c})}_{ds}(\,\vec{Q}_1,\vec{Q}_2\,)\ 
\delta_{f_1\,d}\, \delta_{f_2\,s}
-\widetilde{\phi}^{\,(\Xi^0_{c})}_{sd}(\,\vec{Q}_1,\vec{Q}_2\,)\ 
\delta_{f_1\,s}\, \delta_{f_2\,d}]\, \delta_{f_3\,c} \nonumber\\
&&\hspace{2cm} \times\   (1/2,1/2,0;s_1,s_2,0)\,\delta_{s_3\,r} \\
\hat{\psi}^{\,(\Xi'^{\, +}_{c},r)}_{\alpha_1\,\alpha_2\,\alpha_3}(\,\vec{Q}_1,\vec{Q}_2\,)
&=&\frac{\varepsilon_{c_1\,c_2\,c_3}}{\sqrt{3!}}
\ \frac1{\sqrt2}\,[\widetilde{\phi}^{\,(\Xi'^{\, +}_{c})}_{us}(\,\vec{Q}_1,\vec{Q}_2\,)\ 
\delta_{f_1\,u}\, \delta_{f_2\,s}
+\widetilde{\phi}^{\,(\Xi'^{\, +}_{c})}_{su}(\,\vec{Q}_1,\vec{Q}_2\,)\ 
\delta_{f_1\,s}\, \delta_{f_2\,u}
]\, \delta_{f_3\,c}\nonumber\\
&&\hspace{2cm} \times\   (1/2,1/2,1;s_1,s_2,s_1+s_2)\ 
(1,1/2,1/2;s_1+s_2,s_3,r)\\ 
\hat{\psi}^{\,(\Xi'^{\, 0}_{c},r)}_{\alpha_1\,\alpha_2\,\alpha_3}(\,\vec{Q}_1,\vec{Q}_2\,)
&=&\frac{\varepsilon_{c_1\,c_2\,c_3}}{\sqrt{3!}}
\ \frac1{\sqrt2}\,[\widetilde{\phi}^{\,(\Xi'^{\, 0}_{c})}_{ds}(\,\vec{Q}_1,\vec{Q}_2\,)\ 
\delta_{f_1\,d}\, \delta_{f_2\,s}
+\widetilde{\phi}^{\,(\Xi'^{\, 0}_{c})}_{sd}(\,\vec{Q}_1,\vec{Q}_2\,)\ 
\delta_{f_1\,s}\, \delta_{f_2\,d}
]\, \delta_{f_3\,c}\nonumber\\ 
&&\hspace{2cm} \times\   (1/2,1/2,1;s_1,s_2,s_1+s_2)\ 
(1,1/2,1/2;s_1+s_2,s_3,r)\\
\hat{\psi}^{\,(\Xi^{*\, +}_{c},r)}_{\alpha_1\,\alpha_2\,\alpha_3}(\,\vec{Q}_1,\vec{Q}_2\,)
&=&\frac{\varepsilon_{c_1\,c_2\,c_3}}{\sqrt{3!}}
\ \frac1{\sqrt2}\,[\widetilde{\phi}^{\,(\Xi^{*\, +}_{c})}_{us}(\,\vec{Q}_1,\vec{Q}_2\,)\ 
\delta_{f_1\,u}\, \delta_{f_2\,s}
+\widetilde{\phi}^{\,(\Xi^{*\, +}_{c})}_{su}(\,\vec{Q}_1,\vec{Q}_2\,)\ 
\delta_{f_1\,s}\, \delta_{f_2\,u}]\, \delta_{f_3\,c}\nonumber\\
&&\hspace{2cm} \times\   (1/2,1/2,1;s_1,s_2,s_1+s_2)\ 
(1,1/2,3/2;s_1+s_2,s_3,r) \\
\hat{\psi}^{\,(\Xi^{*\, 0}_{c},r)}_{\alpha_1\,\alpha_2\,\alpha_3}(\,\vec{Q}_1,\vec{Q}_2\,)
&=&\frac{\varepsilon_{c_1\,c_2\,c_3}}{\sqrt{3!}}
\ \frac1{\sqrt2}\,[\widetilde{\phi}^{\,(\Xi^{*\, 0}_{c})}_{ds}(\,\vec{Q}_1,\vec{Q}_2\,)\ 
\delta_{f_1\,d}\, \delta_{f_2\,s}
+\widetilde{\phi}^{\,(\Xi^{*\, 0}_{c})}_{sd}(\,\vec{Q}_1,\vec{Q}_2\,)\ 
\delta_{f_1\,s}\, \delta_{f_2\,d}]\, \delta_{f_3\,c}\nonumber\\
&&\hspace{2cm} \times\   (1/2,1/2,1;s_1,s_2,s_1+s_2)\ 
(1,1/2,3/2;s_1+s_2,s_3,r) \\
\hat{\psi}^{\,(\Omega^{ 0}_{c},r)}_{\alpha_1\,\alpha_2\,\alpha_3}(\,\vec{Q}_1,\vec{Q}_2\,)
&=&\frac{\varepsilon_{c_1\,c_2\,c_3}}{\sqrt{3!}}
\ \widetilde{\phi}^{\,(\Omega^{ 0}_{c})}(\,\vec{Q}_1,\vec{Q}_2\,)\ 
\delta_{f_1\,s}\, \delta_{f_2\,s}
\, \delta_{f_3\,c}\nonumber\\
&&\hspace{2cm} \times\   (1/2,1/2,1;s_1,s_2,s_1+s_2)\ 
(1,1/2,1/2;s_1+s_2,s_3,r) \\
\hat{\psi}^{\,(\Omega^{*\, 0}_{c},r)}_{\alpha_1\,\alpha_2\,\alpha_3}(\,\vec{Q}_1,\vec{Q}_2\,)
&=&\frac{\varepsilon_{c_1\,c_2\,c_3}}{\sqrt{3!}}
\ \widetilde{\phi}^{\,(\Omega^{*\,  0}_{c})}(\,\vec{Q}_1,\vec{Q}_2\,)\ 
\delta_{f_1\,s}\, \delta_{f_2\,s}
\, \delta_{f_3\,c}\nonumber\\
&&\hspace{2cm} \times\   (1/2,1/2,1;s_1,s_2,s_1+s_2)\ 
(1,1/2,1/2;s_1+s_2,s_3,r) 
\eea
Here, besides the properties above, the relation
$\widetilde{\phi}_{sn}(\,\vec{Q}_1,\vec{Q}_2\,)=
\widetilde{\phi}_{ns}(\,\vec{Q}_2,\vec{Q}_1\,)$, with $n=u,d$, also
applies.

These momentum space wave functions are the Fourier
transform of the corresponding wave functions in coordinate
space. Details on how the latter are evaluated in our model for singly
and doubly heavy baryons can be found in
Refs.~\cite{Albertus:2003sx,Albertus:2006wb}.

The two baryons states $\Xi_c,\,\Xi'_c$ differ just in the spin of the
light degrees of freedom, and thus they could mix under the effect of
the hyperfine interaction between the $c$ quark and any of the light
quarks. We have evaluated this mixing in our model finding it
negligible\footnote{In sharp contrast, spin mixings however play a
fundamental role in the case of the semileptonic~\cite{Roberts:2008wq,
Albertus:2009ww} and electromagnetic~\cite{Albertus:2010hi} decays of
the $bc$ baryons.}. Using the AL1 potential, the physical states resulting from the mixing are
$\Xi_c^{(1)}=0.999\ \Xi_c-0.0437 \ \Xi'_c$ and $\Xi_c^{(2)}=0.0437\
\Xi_c+0.999\ \Xi'_c$ , being the mass changes of just $0.2\,$MeV with
respect to the unmixed state case. We neglect this small mixing in our
calculation.
\section{Form factors and weak matrix elements}
\label{app:ffwme}
Taking the initial baryon at rest and $\vec q$ in the positive $Z$ direction 
we define vector and axial matrix elements
\bea
V^\mu_{r\to r'}-A^\mu_{r\to r'}=\big\langle B', r'\ \vec{P}^{\,\prime}=-\vec q\left|\,
\overline \Psi_l(0)\gamma^\mu(1-\gamma_5)\Psi_c(0)
 \right| B, r\ \vec{P}=\vec 0
\big\rangle
\eea
In terms of matrix elements, the different form factors for the spin 1/2-baryon to spin 1/2-baryon 
transitions can be evaluated as
\bea
 F_1&=&-\sqrt{\frac{E'+M'}{2M}}\frac1{|\vec q\,|}V^1_{-1/2\to
1/2}\\
 F_2&=&\frac1{\sqrt{(E'+M')2M}}\bigg(V^0_{1/2\to1/2}+\frac{E'}{|\vec q\,|}V^3_{1/2\to
1/2}+\frac{M'}{|\vec q\,|}V^1_{-1/2\to1/2}\bigg)\\
 F_3&=&-\frac1{\sqrt{(E'+M')2M}}\frac{M'}{|\vec q\,|}\left( V^3_{1/2\to
1/2}-V^1_{-1/2\to
1/2}\right)\\
 G_1&=&\frac1{\sqrt{(E'+M')2M}}A^1_{-1/2\to
1/2}\\
 G_2&=&\sqrt{\frac{E'+M'}{2M}}\frac1{|\vec q\,|}\left(A^0_{1/2\to1/2}
-\frac{M'}{|\vec q\,|}A^1_{-1/2\to
1/2}+\frac{E'}{|\vec q\,|}A^3_{1/2\to1/2}\right)\\
 G_3&=&-\sqrt{\frac{E'+M'}{2M}}\frac{M'}{|\vec q\,|^2}\left( A^3_{1/2\to
1/2}-A^1_{-1/2\to
1/2}\right)
\eea
For the spin 1/2-baryon to spin 3/2-baryon case the relations between 
form factors and weak matrix elements are
\bea
C_3^V&=&\frac{M'}{|\vec q\,|}\frac{1}{\sqrt{(E'+M')2M}}\frac{1}{\sqrt2}\
\left(\,V^1_{1/2\to 3/2}+\sqrt3\, V^1_{1/2\to -1/2}\right)\\
C_4^V&=&\frac{1}{|\vec q\,|^3}\sqrt\frac{E'+M'}{2M}\frac{1}{\sqrt2}\
\left(-\sqrt3 MM'\,V^3_{1/2\to1/2}+M(-2E'+M')\,V^1_{1/2\to 3/2}+
\sqrt3 MM'\,V^1_{1/2\to -1/2}\right)\\
C_5^V&=&\frac{1}{|\vec q\,|^3}\sqrt\frac{E'+M'}{2M}\frac{1}{\sqrt2}\
\left(\sqrt3 |\vec q\,|M'\,V^0_{1/2\to1/2}+\sqrt3 E'M'\,V^3_{1/2\to 1/2}+
M'^2\,V^1_{1/2\to 3/2}-\sqrt3 M'^2\,V^1_{1/2\to -1/2}\right)\\
C_6^V&=&\frac{1}{|\vec q\,|^3}\sqrt\frac{E'+M'}{2M}\frac{1}{\sqrt2}\
\left(-\sqrt3 |\vec q\,|M'\frac{M-E'}{M}\,V^0_{1/2\to1/2}+\sqrt3 |\vec
q\,|^2\frac{M'}{M}\, V^3_{1/2\to 1/2}\right)\\
C_3^A&=&-\frac{M'}{|\vec q\,|^2}\sqrt\frac{E'+M'}{2M}\frac{1}{\sqrt2}\
\left(\,A^1_{1/2\to 3/2}+\sqrt3\, A^1_{1/2\to -1/2}\right)\\
C_4^A&=&-\frac{M'}{|\vec q\,|}\frac{1}{\sqrt{(E'+M')2M}}\sqrt\frac32
\left(\,A^0_{1/2\to1/2}+\frac{E'-M}{|\vec q\,|}\,A^3_{1/2\to
1/2}\right)\nonumber\\
&&+
\frac{1}{M|\vec q\,|^2}\frac{1}{\sqrt{(E'+M')2M}}\frac{1}{\sqrt2}
\left(\, \left(2M^2(E'+M')-MM'(M+M')\right)\,A^1_{1/2\to 3/2}+\sqrt3 MM'(M+M')\, A^1_{1/2\to
-1/2}\right)\nonumber\\\\
C_5^A&=&\frac{M'}{|\vec
q\,|}\frac{1}{\sqrt{(E'+M')2M}}\frac{ME'-M'^2}{M^2}\sqrt\frac32\
\left(\,A^0_{1/2\to1/2}+\frac{E'-M}{|\vec q\,|}\,A^3_{1/2\to
1/2}\right)\nonumber\\
&&+
\frac{1}{M|\vec q\,|^2}\frac{1}{\sqrt{(E'+M')2M}}
\frac{M'^2}{M}\left(2M(E'+M')-(M+M')^2\right)
\frac{1}{\sqrt2}
\left(\,A^1_{1/2\to 3/2}-\sqrt3\, A^1_{1/2\to
-1/2}\right)
\eea
\bea
C_6^A&=&\frac{M'}{|\vec
q\,|}\frac{1}{\sqrt{(E'+M')2M}}\sqrt\frac32
\left(\,A^0_{1/2\to1/2}+\frac{E'}{|\vec q\,|}\,A^3_{1/2\to 1/2}\right)+
\frac{M'^2}{|\vec q\,|^2}\frac{1}{\sqrt{(E'+M')2M}}
\frac{1}{\sqrt2}
\left(\,A^1_{1/2\to 3/2}-\sqrt3\, A^1_{1/2\to
-1/2}\right)\nonumber\\
\eea
For this latter case, 1/2-baryon to 3/2-baryon transitions,  the 
following restrictions are observed
\bea
V^0_{1/2\to1/2}=V^3_{1/2\to1/2}=0\\
V^1_{1/2\to-1/2}=V^1_{-1/2\to1/2}\ \ ,\ \ 
V^1_{1/2\to3/2}=\sqrt3\ V^1_{-1/2\to1/2}\\
A^1_{1/2\to-1/2}=-A^1_{-1/2\to1/2}\ \ , \ \
A^1_{1/2\to3/2}=\sqrt3\ A^1_{-1/2\to1/2}
\eea
so that
\bea
C_3^V&=&\frac{M'}{|\vec q\,|}\frac{1}{\sqrt{2M(E'+M')}}{\sqrt6}\
V^1_{-1/2\to 1/2}\\
C_4^V&=&-\frac{M}{M'}\ C_3^V\\
C_5^V&=&C_6^V=0
\eea
\bea
C_3^A&=&0\\
C_4^A&=&\frac{1}{\sqrt{(E'+M')2M}}\sqrt\frac32\left[
-\frac{M'}{|\vec q\,|}\left(A^0_{1/2\to1/2}+\frac{E'-M}{|\vec q\,|}\,
A^3_{1/2\to 1/2}\right)+
\frac{2(ME'-M'^2)}{|\vec q\,|^2}A^1_{-1/2\to 1/2}\right]
\\
C_5^A&=&\frac{M'}{|\vec
q\,|}\frac{1}{\sqrt{(E'+M')2M}}\sqrt\frac32\left[\ \ \frac{ME'-M'^2}{M^2}
\left(A^0_{1/2\to1/2}+\frac{E'-M}{|\vec q\,|}A^3_{1/2\to 1/2}\right)\right.\nonumber \\
&&\hspace{4cm}+\left.
\frac{2M'(2ME'-M^2-M'^2)}{M^2|\vec q\,|}\
A^1_{-1/2\to 1/2}\right]\\
C_6^A&=&\frac{M'}{|\vec
q\,|}\frac{1}{\sqrt{(E'+M')2M}}\sqrt\frac32\left(
A^0_{1/2\to1/2}+\frac{E'}{|\vec q\,|}A^3_{1/2\to 1/2}+
\frac{2M'}{|\vec q\,|}\,A^1_{-1/2\to 1/2}\right)
\eea\vspace{.5cm}

The  vector  matrix elements have the general structure
\bea
{ V}^{ 0}_{1/2\to 1/2}&=&V^{(0)}_{SF}
\sqrt{2M}\sqrt{2E'}\int d^3Q_1\int d^3Q_2\ 
 \left[\tilde\phi^{(B')}(\vec Q_1-\frac{m_c+m_{l'}}{\overline{M'}}\,\vec q,-\vec Q_1-\vec
 Q_2+\frac{m_{l'}}{\overline{M'}}\,\vec q\,)\right]^*\tilde\phi^{(B)}(\vec Q_1,\vec Q_2)\nonumber\\
&&\times\, \sqrt\frac{(E_l(|\vec{Q}_1-\vec{q}\,
|)+m_l)(E_c(|\vec{Q}_1|)+m_c)}{2E_l(|\vec{Q}_1-\vec{q}\, |)2E_c(|\vec{Q}_1|)}
\left(1+\frac{|\vec{Q}_1|^2-|\vec{q}\, |Q_1^z}{(E_l(|\vec{Q}_1-\vec{q}\, |)+m_l)(E_c(|\vec{Q}_1|)+m_c)}
\right)
\eea
\bea
{ V}^{ 3}_{1/2\to 1/2}&=&V^{(3)}_{SF}
\sqrt{2M}\sqrt{2E'}\int d^3Q_1\int d^3Q_2\ 
 \left[\tilde\phi^{(
 B')}(\vec Q_1-\frac{m_c+m_{l'}}{\overline{M'}}\,\vec q,-\vec Q_1-\vec
 Q_2+\frac{m_{l'}}{\overline{M'}}\,\vec q\,)\right]^*\tilde\phi^{(B)}(\vec Q_1,\vec Q_2)\nonumber\\
&&\times\, \sqrt\frac{(E_l(|\vec{Q}_1-\vec{q}\,
|)+m_l)(E_c(|\vec{Q}_1|)+m_c)}{2E_l(|\vec{Q}_1-\vec{q}\, |)2E_c(|\vec{Q}_1|)}
\left(\frac{Q_1^z}{E_c(|\vec{Q}_1|)+m_c}+\frac{Q_1^z-|\vec{q}\,|}
{E_l(|\vec{Q}_1-\vec{q}\, |)+m_l}
\right)
\eea
\bea
{ V}^{ 1}_{-1/2\to 1/2}&=&V^{(1)}_{SF}
\sqrt{2M}\sqrt{2E'}\int d^3Q_1\int d^3Q_2\ 
 \left[\tilde\phi^{(
 B')}(\vec Q_1-\frac{m_c+m_{l'}}{\overline{M'}}\,\vec q,-\vec Q_1-\vec
 Q_2+\frac{m_{l'}}{\overline{M'}}\,\vec q\,)\right]^*\tilde\phi^{(B)}(\vec Q_1,\vec Q_2)\nonumber\\
&&\times\, \sqrt\frac{(E_l(|\vec{Q}_1-\vec{q}\, |)+m_l)
(E_c(|\vec{Q}_1|)+m_c)}{2E_l(|\vec{Q}_1-\vec{q}\, |)2E_c(|\vec{Q}_1|)}
\nonumber\\
&&\times\, \frac{|\vec{q}\,|(E_c(|\vec{Q}_1|)+m_c)-[E_c(|\vec{Q}_1|)+m_c-
E_l(|\vec{Q}_1-\vec{q}\, |)-m_l]\,Q_1^z}{(E_l(|\vec{Q}_1-\vec{q}\, |)+m_l)(E_c(|\vec{Q}_1|)+m_c)}
\eea
Here we have a $c\to l$ transition at the quark level, while $l'$ is
the light quark originally present in the initial baryon. The
$V^{(j)}_{SF}$ depend on the flavour and spin structure of the baryons
involved. Their values for the different transitions appear in
Table~\ref{tab:SVAF}. When the final baryon has just one $s$ quark
then $\tilde\phi^{( B')}$ should be interpreted as $\tilde\phi^{(
B')}_{sn}$ or $\tilde\phi^{( B')}_{ds}$,  for the case of 
 $c\to s$ or $c\to d$ transitions, respectively.
\begin{table}
\begin{tabular}{lcccccc}\hline
&\ \ $V^{(0)}_{SF}$\ \ &\ \ $V^{(3)}_{SF}$\ \ &\ \ $V^{(1)}_{SF}$\ \ &\ \ $A^{(0)}_{SF}$\ \ &\ \ $A^{(3)}_{SF}$\ \ &\ \ $A^{(1)}_{SF}$\ \ 
\vspace{.1cm}\\\hline
$\Xi_{cc}^{++}\to\Xi^+_c$
 &$\frac{\sqrt3}{\sqrt2}$&$\frac{\sqrt3}{\sqrt2}$&$\frac{-1}{\sqrt6}$
 &$\frac{1}{\sqrt6}$&$\frac{1}{\sqrt6}$&$\frac{1}{\sqrt6}$\\
 
$\Xi_{cc}^{+\hspace{.18cm}}\to\Xi^0_c$
 &$\frac{\sqrt3}{\sqrt2}$&$\frac{\sqrt3}{\sqrt2}$&$\frac{-1}{\sqrt6}$
 &$\frac{1}{\sqrt6}$&$\frac{1}{\sqrt6}$&$\frac{1}{\sqrt6}$\\
 
$\Xi_{cc}^{++}\to\Xi'^+_c$
 &$\frac1{\sqrt2}$&$\frac1{\sqrt2}$&$\frac{-5\sqrt2}{6}$
 &$\frac{5\sqrt2}6$&$\frac{5\sqrt2}6$&$\frac{5\sqrt2}6$\\
 
$\Xi_{cc}^{+\hspace{.18cm}}\to\Xi'^0_c$
 &$\frac1{\sqrt2}$&$\frac1{\sqrt2}$&$\frac{-5\sqrt2}{6}$
 &$\frac{5\sqrt2}6$&$\frac{5\sqrt2}6$&$\frac{5\sqrt2}6$\\
 
$\Xi_{cc}^{++}\to\Xi^{*\,+}_c$
 &$0$&$0$&$\frac{-1}3$
 &$\frac{-2}3$&$\frac{-2}3$&$\frac13$\\
 
$\Xi_{cc}^{+\hspace{.18cm}}\to\Xi^{*\,0}_c$
 &$0$&$0$&$\frac{-1}3$ 
 &$\frac{-2}3$&$\frac{-2}3$&$\frac13$\\

$\Xi_{cc}^{++}\to\Lambda^+_c$
 &$\frac{\sqrt3}{\sqrt2}$&$\frac{\sqrt3}{\sqrt2}$&$\frac{-1}{\sqrt6}$
 &$\frac1{\sqrt6}$&$\frac1{\sqrt6}$&$\frac1{\sqrt6}$\\
 
$\Xi_{cc}^{++}\to\Sigma^+_c$
 &$\frac1{\sqrt2}$&$\frac1{\sqrt2}$&$\frac{-5\sqrt2}6$
 &$\frac{5\sqrt2}6$&$\frac{5\sqrt2}6$&$\frac{5\sqrt2}6$\\
 
$\Xi_{cc}^{+\hspace{.18cm} }\to\Sigma^0_c$
 &$1$&$1$&$\frac{-5}3$
 &$\frac53$&$\frac53$&$\frac{5}3$\\
 
$\Xi_{cc}^{++}\to\Sigma^{*\,+}_c$
 &$0$&$0$&$\frac{-1}{3}$
 &$\frac{-2}3$&$\frac{-2}3$&$\frac{1}{3}$\\
 
$\Xi_{cc}^{+\hspace{.18cm}}\to\Sigma^{*\,0}_c$
 &$0$&$0$&$\frac{-\sqrt2}{3}$
 &$\frac{-2\sqrt2}{3}$&$\frac{-2\sqrt2}{3}$&$\frac{\sqrt2}{3}$\\

$\Omega_{cc}^{+\hspace{.18cm}}\to\Omega^{0}_c$
 &$1$&$1$&$\frac{-5}3$
 &$\frac53$&$\frac53$&$\frac53$\\
 
$\Omega_{cc}^{+\hspace{.18cm}}\to\Omega^{*\,0}_c$
 &$0$&$0$&$\frac{-\sqrt2}3$
 &$\frac{-2\sqrt2}3$&$\frac{-2\sqrt2}3$&$\frac{\sqrt2}3$\\
$\Omega_{cc}^{+\hspace{.18cm}}\to\Xi^{0}_c$
 &$\frac{-\sqrt3}{\sqrt2}$&$\frac{-\sqrt3}{\sqrt2}$&$\frac{1}{\sqrt6}$
 &$\frac{-1}{\sqrt6}$&$\frac{-1}{\sqrt6}$&$\frac{-1}{\sqrt6}$\\
 
$\Omega_{cc}^{+\hspace{.18cm}}\to\Xi'^{0}_c$
 &$\frac1{\sqrt2}$&$\frac1{\sqrt2}$&$\frac{-5\sqrt2}{6}$
 &$\frac{5\sqrt2}{6}$&$\frac{5\sqrt2}{6}$&$\frac{5\sqrt2}{6}$\\
 
$\Omega_{cc}^{+\hspace{.18cm}}\to\Xi^{*\,0}_c$
 &$0$&$0$&$\frac{-1}3$
 &$\frac{-2}3$&$\frac{-2}3$&$\frac13$\\
 
\hline
\end{tabular}
\caption{$V^{(j)}_{SF}$ and $A^{(j)}_{SF}$ spin-flavour factors.}
\label{tab:SVAF}
\end{table}

Similarly, for the axial matrix elements we have
\bea
{ A}^{ 0}_{1/2\to 1/2}&=&A^{(0)}_{SF}
\sqrt{2M}\sqrt{2E'}\int d^3Q_1\int d^3Q_2\ 
 \left[\tilde\phi^{(
 B')}(\vec Q_1-\frac{m_c+m_{l'}}{\overline{M'}}\,\vec q,-\vec Q_1-\vec
 Q_2+\frac{m_{l'}}{\overline{M'}}\,\vec q\,)\right]^*\tilde\phi^{(B)}(\vec Q_1,\vec Q_2)\nonumber\\
&&\times\,  \sqrt\frac{(E_l(|\vec{Q}_1-\vec{q}\, |)+m_l)
(E_c(|\vec{Q}_1|)+m_c)}{2E_l(|\vec{Q}_1-\vec{q}\, |)2E_c(|\vec{Q}_1|)}
\left(\frac{{Q}_1^z}{E_c(|\vec{Q}_1|)+m_c}
+
\frac{{Q}_1^z-|\vec{q}\, |}{E_l(|\vec{Q}_1-\vec{q}\, |)+m_l}
\right)
\eea
\bea
{ A}^{ 3}_{1/2\to 1/2}&=&A^{(3)}_{SF}
\sqrt{2M}\sqrt{2E'}\int d^3Q_1\int d^3Q_2\ 
 \left[\tilde\phi^{(
 B')}(\vec Q_1-\frac{m_c+m_{l'}}{\overline{M'}}\,\vec q,-\vec Q_1-\vec
 Q_2+\frac{m_{l'}}{\overline{M'}}\,\vec q\,)\right]^*\tilde\phi^{(B)}(\vec Q_1,\vec Q_2)\nonumber\\
&&\times\,  \sqrt\frac{(E_l(|\vec{Q}_1-\vec{q}\, |)+m_l)
(E_c(|\vec{Q}_1|)+m_c)}{2E_n(|\vec{Q}_1-\vec{q}\, |)2E_c(|\vec{Q}_1|)}
\left(1-\frac{|\vec{Q}_1|^2-|\vec{q}\, |Q_1^z-2Q_1^z(Q_1^z-|\vec q\,|)}
{(E_l(|\vec{Q}_1-\vec{q}\, |)+m_l)
(E_c(|\vec{Q}_1|)+m_c)}
\right)
\eea
\bea
{ A}^{ 1}_{-1/2\to 1/2}&=&A^{(1)}_{SF}
\sqrt{2M}\sqrt{2E'}\int d^3Q_1\int d^3Q_2\ 
 \left[\tilde\phi^{(
 B')}(\vec Q_1-\frac{m_c+m_{l'}}{\overline{M'}}\,\vec q,-\vec Q_1-\vec
 Q_2+\frac{m_{l'}}{\overline{M'}}\,\vec q\,)\right]^*\tilde\phi^{(B)}(\vec Q_1,\vec Q_2)\nonumber\\
&&\times\,  \sqrt\frac{(E_l(|\vec{Q}_1-\vec{q}\, |)+m_l)
(E_c(|\vec{Q}_1|)+m_c)}{2E_l(|\vec{Q}_1-\vec{q}\, |)2E_c(|\vec{Q}_1|)}
\ \left(1-\frac{|\vec{Q}_1|^2-|\vec{q}\, |Q_1^z-2Q_1^x(Q_1^x-iQ_1^y)}
{(E_l(|\vec{Q}_1-\vec{q}\, |)+m_l)
(E_c(|\vec{Q}_1|)+m_c)}
\right)
\eea
where the $A^{(j)}_{SF}$ axial spin-flavour factors can be found in 
Table~\ref{tab:SVAF}. Note that due to symmetry properties the integral in
$2Q_1^xQ_1^x$ in ${ A}^{ 1}_{-1/2\to 1/2}$ es equivalent to an integral in
$|\vec Q_1|^2-(Q_1^z)^2$, while the integral
in $2Q_1^xQ_1^y$ is identically zero.


\begin{thebibliography}{blabla}
\bibitem{mattson02} M. Mattson {\it et al.} (SELEX Collaboration), Phys. Rev.
Lett. 89, 112001 (2002).
%
\bibitem{ochera05} A. Ocherashvili {\it et al.} (SELEX Collaboration), Phys.
Lett. B 628 (2005) 18.
\bibitem{focus03} S.P. Ratti (FOCUS Collaboration), Nuc. Phys. B (Proc.
Suppl.) 115, 33 (2003). See also\\
 \hbox{www-focus.fnal.gov/xicc/xicc\_focus.html}.
%
\bibitem{babar06} B. Aubert {\it et al.} ({\sl BABAR} Collaboration), Phys.
Rev. D 74, 011103 (2006).
%
\bibitem{lesiak06} R. Chistov (BELLE Collaboration), Phys. Rev. Lett. 97,
162001 (2006).
%
\bibitem{pdg10} K. Nakamura et al. (Particle Data Group), J. Phys. G 37, 
075021 (2010).
%
%
\bibitem{Gershtein:1998sx}
  S.~S.~Gershtein, V.~V.~Kiselev, A.~K.~Likhoded and A.~I.~Onishchenko,
  Mod.\ Phys.\ Lett.\  A {\bf 14}, 135 (1999).
%
\bibitem{Kiselev:1999zj}
  V.~V.~Kiselev and A.~I.~Onishchenko,
  Nucl.\ Phys.\  B {\bf 581}, 432 (2000).
%
\bibitem{Itoh:2000um}
  C.~Itoh, T.~Minamikawa, K.~Miura and T.~Watanabe,
  Phys.\ Rev.\  D {\bf 61}, 057502 (2000).
%
\bibitem{Matrasulov:2000us}
  D.~U.~Matrasulov, M.~M.~Musakhanov and T.~Morii,
  Phys.\ Rev.\  C {\bf 61}, 045204 (2000).
%
\bibitem{Gershtein:2000nx}
  S.~S.~Gershtein, V.~V.~Kiselev, A.~K.~Likhoded and A.~I.~Onishchenko,
  Phys.\ Rev.\  D {\bf 62}, 054021 (2000).
%
%
\bibitem{Kiselev:2000jb}
  V.~V.~Kiselev and A.~E.~Kovalsky,
  Phys.\ Rev.\  D {\bf 64}, 014002 (2001).
  %
  %
\bibitem{lewis01} R. Lewis, N. Mathur, and R.M. Woloshyn, 
Phys. Rev D 64, 094509 (2001).
%
\bibitem{Kiselev:2002iy}
  V.~V.~Kiselev, A.~K.~Likhoded, O.~N.~Pakhomova and V.~A.~Saleev,
  Phys.\ Rev.\  D {\bf 66}, 034030 (2002).
%
%
\bibitem{Ebert:2002ig}
  D.~Ebert, R.~N.~Faustov, V.~O.~Galkin and A.~P.~Martynenko,
  Phys.\ Rev.\  D {\bf 66}, 014008 (2002).
%
\bibitem{Mathur:2002ce}
  N.~Mathur, R.~Lewis and R.~M.~Woloshyn,
  Phys.\ Rev.\  D {\bf 66}, 014502 (2002)
%
\bibitem{Flynn:2003vz}
  J.~M.~Flynn, F.~Mescia and A.~S.~B.~Tariq  [UKQCD Collaboration],
  JHEP {\bf 0307}, 066 (2003).
%
\bibitem{Vijande:2004at}
  J.~Vijande, H.~Garcilazo, A.~Valcarce and F.~Fernandez,
  Phys.\ Rev.\  D {\bf 70}, 054022 (2004).
%
%
\bibitem{Ebert:2005ip}
  D.~Ebert, R.~N.~Faustov, V.~O.~Galkin and A.~P.~Martynenko,
  Phys.\ Atom.\ Nucl.\  {\bf 68} (2005) 784
  [Yad.\ Fiz.\  {\bf 68} (2005) 817].
\bibitem{Mehen:2006vv}
  T.~Mehen and B.~C.~Tiburzi,
  Phys.\ Rev.\  D {\bf 74}, 054505 (2006).
%
\bibitem{Albertus:2006wb}
  C.~Albertus, E.~Hernandez, J.~Nieves and J.~M.~Verde-Velasco,
  Eur.\ Phys.\ J.\  A {\bf 31}, 691 (2007); erratum ibid. Eur. Phys. J. A 36,
  119  (2008).

\bibitem{Martynenko:2007je}
  A.~P.~Martynenko,
  Phys.\ Lett.\  B {\bf 663}, 317 (2008).
\bibitem{Zhang:2008rt}
  J.~R.~Zhang and M.~Q.~Huang,
  Phys.\ Rev.\  D {\bf 78}, 094007 (2008).
%
\bibitem{Giannuzzi:2009gh}
  F.~Giannuzzi,
  Phys.\ Rev.\  D {\bf 79}, 094002 (2009).
 %
\bibitem{Albertus:2009ww}
  C.~Albertus, E.~Hernandez and J.~Nieves,
  Phys.\ Lett.\  B {\bf 683}, 21 (2010).
 %
\bibitem{Liu:2009jc}
  L.~Liu, H.~W.~Lin, K.~Orginos and A.~Walker-Loud,
  Phys.\ Rev.\  D {\bf 81}, 094505 (2010).

\bibitem{Narison:2010py}
  S.~Narison and R.~Albuquerque,
  Phys.\ Lett.\  B {\bf 694}, 217 (2010).
%
\bibitem{weng} M.-H. Weng,  X.-H. Guo,  A.W. Thomas,  Phys. Rev. D 83, 056006
(2011). 
%
%
%
\bibitem{Kiselev:1998sy}
  V.~V.~Kiselev, A.~K.~Likhoded and A.~I.~Onishchenko,
  Phys.\ Rev.\  D {\bf 60}, 014007 (1999).
%
%
\bibitem{Guberina:1999mx}
  B.~Guberina, B.~Melic and H.~Stefancic,
  Eur.\ Phys.\ J.\  C {\bf 9}, 213 (1999); erratum ibid.
  Eur.\ Phys.\ J.\  C {\bf 13}, 551 (2000).
%
\bibitem{Kiselev:2001fw}
  V.~V.~Kiselev and A.~K.~Likhoded,
  Phys.\ Usp.\  {\bf 45}, 455 (2002)
  [Usp.\ Fiz.\ Nauk {\bf 172}, 497 (2002)]
  [arXiv:hep-ph/0103169].
See also A.I. Onishchenko, hep-ph/9912425; A.I. Onishchenko, hep-ph/0006271;
A.I. Onishchenko, hep-ph/0006295.
\bibitem{Chang:2007xa}
  C.~H.~Chang, T.~Li, X.~Q.~Li and Y.~M.~Wang,
  Commun.\ Theor.\ Phys.\  {\bf 49}, 993 (2008).
%

\bibitem{Faessler:2001mr}
  A.~Faessler, T.~Gutsche, M.~A.~Ivanov, J.~G.~Korner and V.~E.~Lyubovitskij,
  Phys.\ Lett.\  B {\bf 518}, 55 (2001).

\bibitem{Albertus:2003sx}
  C.~Albertus, J.~E.~Amaro, E.~Hernandez and J.~Nieves,
  Nucl.\ Phys.\  A {\bf 740}, 333 (2004).
%

%
\bibitem{semay94} C. Semay, and B. Silvestre-Brac, Z. Phys. C 61, 271 (1994).
\bibitem{silvestre96} B. Silvestre-Brac, Few-Body Systems  20, 1 (1996).
%
%

%
\bibitem{Llewellyn Smith:1971zm}
  C.~H.~Llewellyn Smith,
  Phys.\ Rept.\  {\bf 3}, 261 (1972).

%
\bibitem{hqs1} S. Nussinov and W. Wetzel, Phys. Rev. D 36, 130 (1987).
\bibitem{hqs2} M.A. Shifman and M.B. Voloshin, Sov. J.
Nucl. Phys. 45, 292 (1987) (Yad. Fiz. 45, 463 (1987)) .
\bibitem{hqs3} H.D. Politzer and M.B. Wise, Phys. Lett. B 206, 681 (1988); 
208, 504 (1988).
\bibitem{hqs4} N. Isgur and M.B. Wise, Phys. Lett. B 232, 113 (1989); 
237, 527 (1990).  
\bibitem{thacker91} B.A. Thacker and G.P. Lepage, Phys. Rev. D43, 196 (1991).
\bibitem{Falk:1990yz} A. F. Falk, H. Georgi, B. Grinstein, and M. B. Wise, Nucl.
Phys. B343, 1 (1990).
\bibitem{MWbook} A.V. Manohar and M.B. Wise, {\it Heavy Quark Physics}
  (Cambridge University Press, Cambridge, England, 2000), ISBN 0-521-64241-8.
\bibitem{Flynn:2007qt}
  J.M. Flynn and J. Nieves,  Phys. Rev. D76, 017502 (2007); erratum
  ibid. Phys. Rev. D77, 099901 (2008).
\bibitem{Hernandez:2007qv}
  E.~Hernandez, J.~Nieves, J.~M.~Verde-Velasco,
  Phys.\ Lett.\  B663, 234   (2008).

\bibitem{BD81} R. K. Bhaduri, L.E. Cohler, Y. Nogami, Nuovo Cim. 
A 65, 376 (1981).


\bibitem{Isgur:1989qw} 
N.~Isgur, M.~B.~Wise, 
 Phys.\ Rev.\  D41, 151 (1990).
  


\bibitem{Albertus:2005ud}
  C.~Albertus, J.~M.~Flynn, E.~Hernandez, J.~Nieves, J.~M.~Verde-Velasco,
  Phys.\ Rev.\  D72, 033002 (2005).

\bibitem{Albertus:2004wj}
  C.~Albertus, E.~Hernandez and J.~Nieves,
  Phys.\ Rev.\  D71, 014012 (2005).

\bibitem{HB-Lattice}  K.C. Bowler et al. [UKQCD Collaboration],
  Phys. Rev.  D57 (1998) 6948
%
%
\bibitem{Roberts:2008wq}
  W.~Roberts, M.~Pervin,
  Int.\ J.\ Mod.\ Phys.\  A24, 2401 (2009).


\bibitem{Albertus:2010hi}
  C.~Albertus, E.~Hernandez, J.~Nieves,
  Phys.\ Lett.\  B690,  265 (2010).

%

%
%
%
%
%
%
%
%
%

%


%


\end{thebibliography}
\end{document}